\documentclass[aps,pra, twocolumn, noshowpacs, floatfix]{revtex4}
\pdfoutput=1
\usepackage{wrapfig}
\usepackage{amsfonts}
\usepackage{amssymb}
\usepackage{graphicx}
\usepackage{amsmath}
\usepackage{braket}
\usepackage[english]{babel}
\usepackage[mathscr]{euscript}
\usepackage{color}
\usepackage{epstopdf}
\usepackage{footnote}
\usepackage{isotope}
\usepackage{ulem}
\usepackage{float}
\usepackage{soul}
\usepackage{xcolor}
\sethlcolor{yellow!50}
\usepackage[T1]{fontenc}
\usepackage[utf8]{inputenc}   

\usepackage[export]{adjustbox}
\allowdisplaybreaks

\begin{document}
\title{A Davydov Ansatz approach to accurate system-bath dynamics in the presence of multiple baths with distinct temperatures}

\author{Chenlin Ma$^{1}$, Fulu Zheng$^{2}$\footnote{Electronic address:~\url{zhengfulu@nbu.edu.cn}}, Kewei Sun$^{3}$, Lu Wang$^{4}$, and Yang Zhao$^{1}$\footnote{Corresponding author. Electronic address:~\url{YZhao@ntu.edu.sg}}}

\affiliation{$^{1}$\mbox{School of Materials Science and Engineering, Nanyang Technological University, Singapore 639798, Singapore}\\
$^{2}$\mbox{Department of Physics , Ningbo University, Ningbo 315211, China}\\
$^{3}$\mbox{School of Science, Hangzhou Dianzi University, Hangzhou 310018, China}\\
$^{4}$\mbox{School of Science, Inner Mongolia University of Science and Technology, Inner Mongolia 014010, China} \\
}

\begin{abstract}
We perform benchmark simulations using the time-dependent variational approach with the multiple Davydov Ansatz (mDA) to study real-time nonequilibrium dynamics in a single qubit model coupled to two thermal baths with distinct temperatures. A broad region of the parameter space has been investigated, accompanied by a detailed analysis of the convergence behavior of the mDA method. Additionally, we have compared our mD2 results to those from two widely adopted, numerically ``exact'' techniques: the methods of hierarchical equations of motion (HEOM) and the quasi-adiabatic path integral (QUAPI).
It is found that the mDA approach in combination with thermal field dynamics  yields numerically accurate, convergent results in nearly all regions of the parameter space examined, including those that pose serious challenges for QUAPI and HEOM.
Our results reveal that mDA offers a highly adaptable framework capable of capturing long-time dynamics, even in challenging regimes where other methods face limitations. These findings underscore the potential of mDA as a versatile tool for exploring quantum thermodynamics, energy transfer processes, and non-equilibrium quantum systems.

\end{abstract}

\date{\today}
\maketitle

\section{Introduction}

The study of open quantum systems is essential for understanding the complex interactions between quantum systems and their surrounding environments. In real-world scenarios, no quantum system exists in complete isolation, and its interaction with external baths plays a crucial role in determining its dynamics and transport properties.
Accurately modeling quantum dissipation-where energy is exchanged between a system and its environment-poses a significant challenge in theoretical and computational physics and chemistry \cite{breuer2002theory,rivas2012open,Weiss1999,redf,lindblad1998brownian,vacchini2000completely,tameshtit1996positive,gao1997dissipative,Lidar1998,Lidar1999}.
Central to the problem of quantum dissipation is the question of compatibility between quantum mechanics and markovian motion, as manifested by the Lindblad exclusion principle of density-matrix positivity, translational invariance, and approach to canonical equilibrium.  The difficulty lies in the failure of a Hamiltonian description of those systems, and in the elusiveness of a quantization procedure that ensues.
Traditional approaches such as the Redfield theory \cite{redf} the Lindblad master equation \cite{Lidar1998,Zhao911} offer a practical framework under weak system-bath coupling and Markovian assumptions, but these approximations break down when non-Markovian effects become significant or when strong coupling dominates the dynamics. In such cases, an explicit treatment of environmental degrees of freedom is required, and conventional analytical approaches often prove inadequate.

\begin{figure}[htbp]
\vspace{0.5cm}
  \centering
  \includegraphics[width=0.4\textwidth]{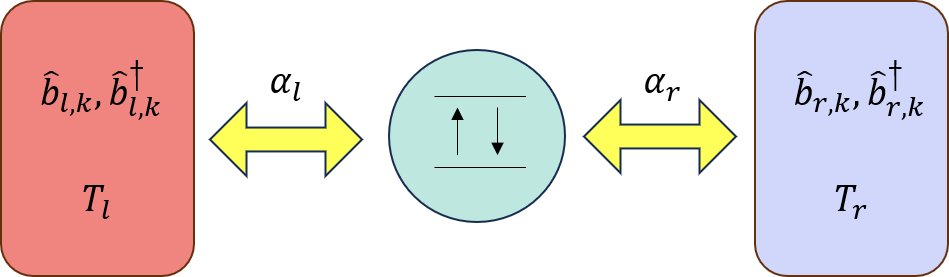}
  \caption{Schematic of the two-bath spin-boson model. A central two-level system is simultaneously coupled to two independent bosonic reservoirs of temperatures $T_l$ and $T_r$, and the $k$th reservoir mode is characterized by annihilation and creation operators $\hat b_{\nu,k}$ and $\hat b_{\nu,k}^\dagger$ ($\nu=l,r$), and the bidirectional arrows denote the system-bath coupling.}
  \label{fig:my_label10}
\end{figure}

 Recent advances in nanoscale physics, condensed matter systems, and experimental techniques have spurred significant interest in quantum thermodynamics and energy transport mechanisms \cite{rivas2014quantum,breuer2016colloquium,de2017dynamics}. As a toy model of non-equilibrium energy transport, the two-bath spin-boson model comprises a single two-level system coupled simultaneously to two independent bosonic reservoirs held at different temperatures, as shown in Fig.~1. Conventional perturbative approaches, such as Bloch-Redfield theory, fail to capture intricate non-equilibrium effects in strongly correlated systems \cite{breuer2002theory}.
To address these challenges, a variety of numerically accurate approaches have been developed. Notably, the quasi-adiabatic path integral (QUAPI) and the hierarchical equations of motion (HEOM) have emerged as robust, non-perturbative frameworks for simulating open quantum systems. QUAPI leverages a path-integral discretization to systematically capture memory and non-Markovian effects \cite{makri1995tensor,makri1995tensor2}, while HEOM constructs a hierarchy of coupled differential equations that encapsulate higher-order system-bath interactions \cite{tanimura2020numerically,tanimura2006stochastic}. Both methods, formulated in terms of the reduced density matrix, are well-suited for computing nonlinear spectra and can be naturally extended to multidimensional spectroscopic signals by propagating the reduced density matrix along multiple Liouville space pathways \cite{ikeda2017probing,liang2014simulating}. Nevertheless, significant challenges remain when applying these techniques to multi-reservoir scenarios such as thermal engines or steady-state heat transport. In particular, HEOM's computational cost grows exponentially with the number of auxiliary density operators (and thus with bath temperature, coupling strength, and hierarchy depth), making convergence difficult for low-temperature or structured spectral densities. QUAPI, by contrast, requires explicit discretization of the bath memory kernel over a finite-memory time, which causes an exponential growth in tensor dimensions (and severe Trotter errors if the time step is not made prohibitively small), effectively limiting it to short-time dynamics and simpler spectral profiles \cite{kato2016quantum,kato2015quantum,nicolin2011non}.

In addition to the aforementioned methods, Zhao and co-workers introduced a variational framework that combines multiple Davydov Ans\''{a}tze (mDA) with the time-dependent variational principle to deliver a numerically accurate, yet efficient description of non-Markovian quantum dynamics \cite{zhao2023hierarchy,zhao2022hierarchy}. In mDA, the bosonic sector of the wave function is expanded as a superposition of coherent-state configurations drawn from an overcomplete basis, which accurately captures intricate multi-mode correlations. When integrated with thermal field dynamics (TFD), mDA naturally extends to finite-temperature open systems, seamlessly incorporating thermal fluctuations and non-equilibrium effects. Crucially, its computational cost scales linearly with the number of bosonic modes, ensuring excellent scalability. To date, mDA has been successfully applied in many problems in
chemical physics, condensed-matter theory, cavity QED, nonlinear spectroscopy, and many-body quantum dynamics \cite{shen2024finite,sun2022exciton,shen2024cavity,zheng2021photon,zhang2024dissipative,sun2019monitoring}, and more recently to non-Hermitian Hamiltonians \cite{zhang2024hamiltonian}. These combined advantages render mDA an ideal toolkit for exploring non-equilibrium energy transport in multi-reservoir open quantum systems. In this work, we employ mDA within a time-dependent variational framework to study the two-bath spin-boson model, examining its non-equilibrium dynamics and energy flows across a range of temperature gradients and coupling strengths.
The dynamics are calculated over the full temperature range, while the calculation of steady-state heat current is performed at low temperatures to further prove the feasibility of the mD2 method for steady-state heat transport. Detailed results of the low-temperature steady-state heat current are provided in Supplementary Material, Section 1.
To validate mDA's performance, we benchmark its accuracy, convergence behavior, and computational efficiency against HEOM and QUAPI.
Our results demonstrate that mDA delivers high-fidelity predictions with linear scaling in the number of bosonic modes, even in regimes where HEOM and QUAPI encounter convergence or memory-kernel limitations, thereby underscoring its suitability for quantum thermodynamics investigations in complex open-system settings.

The remainder of this paper is organized as follows. In Section II, we introduce the model describing a two-level system interacting with two heat baths at different temperatures and provide a comprehensive description of our mDA methodology. Section III presents a detailed benchmark of our approach against established methods, highlighting its advantages and limitations. Finally, Section IV summarizes our principal findings and outlines potential future directions.

\section{METHODOLOGY}

\subsection{Model}
To set the stage, we consider the Hamiltonian of a single qubit that is interacting with two radiation baths. This minimal open quantum system model, commonly employed in studies of phenomena such as heat transport and ion transport \cite{nicolin2011non}, is defined by:
\begin{eqnarray}\label{EQ1}
H_{\text{total}} = H_{q} + H_b + H_{I},
\end{eqnarray}
where the qubit Hamiltonian (with $\hbar = 1$) is specified by 
\begin{eqnarray}\label{EQ2}
H_{q} = \frac{\omega_{0}}{2}\,\sigma_{z}.
\end{eqnarray}
Here $\omega_0$ is the energy splitting, which is set to be 1 eV throughout the study, and we set the tunneling parameter $\eta$ to zero. The usual Pauli operators are denoted by $\sigma_z$ and $\sigma_x$.
The left (L) and right (R) baths, along with the interaction between the qubit and each bath, are described as
\begin{equation}\label{EQ3}
    H_b = \sum_{r = {\rm L, R}} \sum_{k = 1}^{N} \omega_{r,k}\,b_{r,k}^\dagger\,b_{r,k},
\end{equation}
and
\begin{equation}\label{EQ4}
    H_{I} = \sigma_{x} \sum_{r = {\rm L, R}} \sum_{k = 1}^{N} \lambda_{r,k}\,\Bigl(b_{r,k}^\dagger + b_{r,k}\Bigr).
\end{equation}
where $b_{r,k}$ ($b_{r,k}^\dagger$) annihilates (creates) a boson in the $k$th mode of frequency $\omega_{r,k}$ in bath $r$. The coefficient $\lambda_{r,k}$ governs the interaction strength between the qubit and the $k$th mode in bath $r$.
The coupling between the qubit and its baths is characterized by a Drude-Lorentz spectral density:
\[
J(\omega)
= \sum_{n}\kappa_{n}^2\,\delta(\omega - \omega_{n})
= 2\,\alpha \,\frac{\omega_{c}\,\omega}{\omega^{2} + \omega_{c}^{2}},
\]
where \(\delta(\cdot)\) denotes the Dirac delta function, $\kappa_{n}$ represents the coupling strength between the system and the n-th mode of the bath, \(\alpha\) is the dimensionless coupling strength, and \(\omega_{c}\) is the inverse bath correlation time. In the present study, we adopt \(\alpha = 0.015\,\mathrm{eV}\) and \(\omega_{c} = 1.5\,\mathrm{eV}\). Accurate discretization of \(J(\omega)\) is imperative, since the low-temperature dynamics of a two-level system are highly sensitive to the spectral form \cite{chen2021simulation}. It is worth noting that the mD2 version used in this manuscript is the most basic, unoptimized form, and while it is capable of handling more sophisticated spectral densities, we deliberately choose a Drude-Lorentz spectral density for simplicity and to facilitate comparisons. This choice is made because the Drude-Lorentz form is the most convenient for the HEOM method, allowing us to compare the methods in their unoptimized versions. The more advanced capabilities of mD2, including handling arbitrary spectral densities, can be easily incorporated, but this is beyond the scope of the present work. As highlighted in \cite{xu2022taming}, recent advancements in HEOM, such as the free pole method and tensor network incorporation, have expanded its applicability to more general spectral densities. In the future, we plan to extend this comparison by applying more advanced versions of all three methods to arbitrary forms of spectral densities, especially in the context of more complex many-body problems, such as the t-J model.

To efficiently sample the bath degrees of freedom, we employ the interpolative decomposition (ID) discretization scheme described in Ref.\,\cite{takahashi2024discretization}, dividing the domain $[0, \omega_{\max}]$ (with $\omega_{\max} = 10\,\omega_c$) into $N_{b}$ effective intervals. This method assigns the coefficients $\kappa_{n}$ in a way that significantly improves sampling efficiency relative to alternative discretization methods (e.g., logarithmic). Such an approach is critical for capturing finite-temperature effects in the mDA framework, as being domenstrated in Appendix~A.

\subsection{Thermal Field Dynamics}

To handle finite-temperature effects, we use the Thermofield Dynamics (TFD) method, which introduces fictitious bath degrees of freedom to represent thermal states. These fictitious degrees of freedom are introduced only in the bath, allowing us to effectively model the temperature effects in the bath while preserving the quantum coherence of the system. This approach is distinct from transforming to Liouville space, as TFD deals with the thermal effects by introducing auxiliary thermal bath degrees of freedom, rather than altering the system’s representation. Through this method, we are able to accurately capture the energy exchange and temperature fluctuations between the system and the bath, especially in regimes of strong coupling and non-Markovian dynamics. We first express the expectation value of any operator \( Q \) in the vibrational space in terms of a ``thermal vacuum state'' \( |0(\beta)\rangle \)\,\cite{suzuki1985thermo,takahashi1996thermo}:
\begin{equation}
\langle Q \rangle_\beta = \mathrm{Tr}\{Q\,\rho_\beta\} = \langle 0(\beta)| Q |0(\beta)\rangle,
\end{equation}
where \( |0(\beta)\rangle \) is the temperature-dependent ``vacuum state,'' defined as\,\cite{takahashi1996thermo}
\begin{equation}
|0(\beta)\rangle = \frac{1}{\sqrt{Z}} \sum_{k} e^{-\beta \hbar \omega_l/2}\, |k,\tilde{k}\rangle,
\end{equation}
in which
$
|I\rangle = \sum_{k} |k,\tilde{k}\rangle,
$
where \( |k\rangle \) is the eigenstate of the vibrational Hamiltonian \( h_g \) with eigenvalue \( \hbar \omega_k \) and \( |\tilde{k}\rangle \) is the eigenstate of the corresponding fictitious Hamiltonian
\[
\tilde{h}_g = \sum_{k} \hbar \omega_k\, \tilde{b}_k^\dagger \tilde{b}_k.
\]
It is convenient to express the thermal vacuum state as\,\cite{takahashi1996thermo}
\begin{equation}
|0(\beta)\rangle = e^{-iG_{r}}|0\rangle_{g},
\end{equation}
where \( |0\rangle_{g} = |00\rangle_{g} \) is the ground state of the augmented vibrational Hamiltonian (i.e., \(h_g\) and \(\tilde{h}_g\)), and \(G_r\) is the Bogoliubov unitary transformation operator for r bath, defined as
\begin{equation}
G_r = \sum_{k}\theta_{rk}\Bigl( b_{rk}^{\dagger}\tilde{b}_{rk}^{\dagger} - \tilde{b}_{rk}\,b_{rk}\Bigr),
\end{equation}
with
\begin{equation}
\theta_{rk} = \operatorname{arctanh}\Bigl(e^{-\beta_r\hbar\omega_{rk}/2}\Bigr).
\end{equation}
We can thus rewrite Hamiltonian via Bogoliubov transformation quite easily, where $H_\theta = e^{iG_R}e^{iG_L}H_{\rm total}e^{-iG_L}e^{-iG_R}$, using the following relations:
\[
e^{i\hat{G_r}}\,\hat{b}_{r,k}\,e^{-i\hat{G_r}}
= \hat{b}_{r,k}\,\cosh\bigl(\theta_{r,k}\bigr)
+ \hat{b}_{r,k}^{\dagger}\,\sinh\bigl(\theta_{r,k}\bigr),
\]
\[
e^{i\hat{G_r}}\,\hat{b}_{r,k}^{\dagger}\,e^{-i\hat{G_r}}
= \hat{b}_{r,k}^{\dagger}\,\cosh\bigl(\theta_{r,k}\bigr)
+ \hat{b}_{r,k}\,\sinh\bigl(\theta_{r,k}\bigr),
\]
\[
e^{\hat{G_r}}\,
\Bigl(\hat{b}_{r,k}^{\dagger}\,\hat{b}_{r,k}
     \;-\;
     \hat{b}_{r,k}\,\hat{b}_{r,k}^{\dagger}\Bigr)\,
e^{-\hat{G_r}}
= \hat{b}_{r,k}^{\dagger}\,\hat{b}_{r,k}
\;-\;
\hat{b}_{r,k}\,\hat{b}_{r,k}^{\dagger}.
\]
And the effective hamitonian $H_\theta$ is finally given by:
\begin{eqnarray}
H_{\theta} &=& \frac{1}{2}\,\sigma_{z}\,\omega_{0} \nonumber \\
&&+ \sum_{k}\omega_{lk}\,b_{lk}^{\dagger}b_{lk}
 -\sum_{k}\omega_{lk}\,\tilde{b}_{lk}^{\dagger}\tilde{b}_{lk} \nonumber\\
&&+ \sum_{k}\omega_{rk}\,b_{rk}^{\dagger}b_{rk}
 -\sum_{k}\omega_{rk}\,\tilde{b}_{rk}^{\dagger}\tilde{b}_{rk} \nonumber\\
&&+ \sigma_{x}\sum_{k}\lambda_{lk}\Bigl[\bigl(b_{lk}^{\dagger}+b_{lk}\bigr)\cosh(\theta_{lk})\nonumber\\
&&+ \bigl(\tilde{b}_{lk}^{\dagger}+\tilde{b}_{lk}\bigr)\sin(\theta_{lk})\Bigr] \nonumber\\
&&+ \sigma_{x}\sum_{k}\lambda_{rk}\Bigl[\bigl(b_{rk}^{\dagger}+b_{rk}\bigr)\cosh(\theta_{rk})\nonumber\\
 &&+ \bigl(\tilde{b}_{rk}^{\dagger}+\tilde{b}_{rk}\bigr)\sin(\theta_{rk})\Bigr].
\label{eq:Heff}
\end{eqnarray}

At time \( t = 0 \), the initial condition \(\rho(0)\) is given by an uncorrelated product state of the qubit and the baths. Due to the finite temperature of the baths, this state must be described as a mixed state, such as:

\begin{equation}
    \rho(0) = \rho(s) \otimes \frac{e^{-H_{r,\text{bath}}\beta_{r}}}{\text{Tr}(e^{-H_{r,\text{bath}}\beta_{r}})} \otimes \frac{e^{-H_{l,\text{bath}}\beta_{l,}}}{\text{Tr}(e^{-H_{l,\text{bath}}\beta_{l}})}
\end{equation}
where \(\rho(s)\) is an arbitrary initial density matrix of the qubit. For a typical molecular system where the electronic energy gap is much larger than the vibrational energies, we can safely assume that the system resides in the electronic ground state \cite{chen2017finite}. The initial density matrix under this approximation can be written as
$
\rho(0) = |g\rangle\langle g| \otimes \rho_{\mathrm{vib}}(0),
$
where \(\rho_{\mathrm{vib}}(0)\) is the initial density matrix for the vibrational degrees of freedom.

\subsection{The multi-D$_2$ Ansatz}

The Schr\"{o}dinger equation governed by Eq.~(\ref{eq:Heff}) cannot be analytically solved except in special cases.
To obtain a numerically accurate solution to the Schr\"{o}dinger equation, in this work we employ the time-dependent variational principle with the multi-D$_2$ (mD2) Ansatz as the trial wave function. The Multi-D2 Ansatz (mD2 ansatz) method represents the wavefunction as a linear combination of coherent states, capturing the interaction between the system and the bath. This method is particularly advantageous in non-equilibrium systems, as it efficiently models the time evolution of the system, especially under strong coupling and non-Markovian dynamics. When combined with Thermofield Dynamics (TFD), it allows us to account for finite-temperature effects while maintaining the quantum properties of the system. This combination makes the mD2 approach highly effective for simulating systems with strong system-bath coupling and thermal effects, ensuring both accuracy and computational efficiency.
A member of the mDA family of time-dependent variational trial states \cite{zhao2023hierarchy}, the mD2 Ansatz of multiplicity \( M \) can be written as follows:
\begin{equation}
    \begin{aligned}
        \left| {\rm D}_{2}^{M}(t) \right\rangle =& \sum_{i=1}^{M} \left[ A_{i} \vert + \rangle + B_{i} \vert - \rangle \right] \exp \left[ \sum_{k=1}^{N} \left( f_{i,k} b_{lk}^{\dagger} - \text{H.c.} \right) \right] \left| 0 \right\rangle\\
        &\times  \exp \left[ \sum_{k=1}^{N} \left( \tilde{f}_{i,k} \tilde{b}_{lk}^{\dagger} - \text{H.c.} \right) \right] \left| \tilde{0} \right\rangle\\
	   &\times \exp \left[ \sum_{k=1}^{N} \left( g_{i,k} b_{rk}^{\dagger} - \text{H.c.}\right) \right] \left| 0^\prime \right\rangle\\
        &\times  \exp \left[ \sum_{k=1}^{N} \left( \tilde{g}_{i,k} \tilde{b}_{rk}^{\dagger} - \text{H.c.}\right) \right] \left| \tilde{0^\prime} \right\rangle\\
    \end{aligned}
\end{equation}
where $\left| 0 \right\rangle$ denotes the vacuum state for the left bath, and $\left| 0^\prime \right\rangle$ denotes the right bath. The states \(\ket{+}\) and \(\ket{-}\) represent the spin states, with amplitudes \( A_i \) and \( B_i \), respectively. The function \( f_{i} \),\( g_{i} \)  represents the displacement of the boson mode in left and right bath, respectively, with its complex conjugate denoted as \( f_{i}^{\ast} \) (\( \tilde{g}_{i}^{\ast} \)). The states \(\ket{0}\) and \(\ket{0^\prime}\) are the vacuum states of the boson modes for the left and right baths, respectively, in real Hilbert space.
$A_i$, $B_i$, $f_i$ ,$\tilde f_i$,$g_i$,and $\tilde g_i$ are called the variational parameters. These variational parameters are be solved via the Euler-Lagrangian equation:
\begin{align}
\frac{d}{d t} \frac{\partial L}{\partial \dot{u}_{i}^{*}}-\frac{\partial L}{\partial u_{i}^{*}}=0, u_{i} \in\left[A_{i}, B_{i},f_{i, k},\tilde{f}_{i,k},g_{i, k},\tilde{g}_{i,k}\right]
\end{align}
with
\begin{eqnarray}
L&=&\frac{i}{2}\left[\langle {\rm D}_{2}^{M}(t)|\frac{\overrightarrow{\partial}}{\partial t}|{\rm D}_{2}^{M}(t)\rangle
-\langle {\rm D}_{2}^{M}(t)|\frac{\overleftarrow{\partial}}{\partial t}|{\rm D}_{2}^{M}(t)\rangle\right]\nonumber\\&&-\langle{\rm D}_{2}^{M}(t)|\hat{H}_{\theta}|{\rm D}_{2}^{M}(t)\rangle.
\end{eqnarray}
The equations of motion (EOMs) for the variational parameters are numerically solved via the 4$^{th}$ order Runge-Kutta method, as shown in Appendix B. Moreover, the expectation value of an arbitrary observable \( Q \) for the system at finite temperatures can be written as:
\begin{eqnarray}
&&\langle Q(t)\rangle = \bra{{\rm D}_{2}^{M}(t)}Q\ket{{\rm D}_{2}^{M}(t)}
\end{eqnarray}
The population inversion is therefore defined as:
\begin{eqnarray}
\langle{\mathcal P}_{z}\rangle = \bra{D_2^M}{\sigma}_{z}\ket{D_2^M}\nonumber \\
= \sum_{m, n}^{M}(A_{j}^{\ast}{A}_{i}-B_{j}^{\ast}{B}_{i})S_{ij}
\end{eqnarray}
where $S_{ij} $ is the Debye-Waller Factor (DWF), which has the form:
\begin{eqnarray}
S_{ij} &=&
\exp\Biggl\{
  \sum_{k}
  \biggl[
    f_{j,k}^*\,f_{i,k}
    \;-\;
    \tfrac{1}{2}\Bigl(|f_{i,k}|^2 + |f_{j,k}|^2\Bigr)
  \biggr]
  \nonumber\\[6pt]
& & \quad
  + \sum_{k}
  \biggl[
    \tilde{f}_{j,k}^*\,\tilde{f}_{i,k}
    \;-\;
    \tfrac{1}{2}\Bigl(|\tilde{f}_{i,k}|^2 + |\tilde{f}_{j,k}|^2\Bigr)
  \biggr]
  \nonumber\\[6pt]
& & \quad
  + \sum_{k}
  \biggl[
    g_{j,k}^*\,g_{i,k}
    \;-\;
    \tfrac{1}{2}\Bigl(|g_{i,k}|^2 + |g_{j,k}|^2\Bigr)
  \biggr]
  \nonumber\\[6pt]
& & \quad
  + \sum_{k}
  \biggl[
    \tilde{g}_{j,k}^*\,\tilde{g}_{i,k}
    \;-\;
    \tfrac{1}{2}\Bigl(|\tilde{g}_{i,k}|^2 + |\tilde{g}_{j,k}|^2\Bigr)
  \biggr]
\Biggr\}. \nonumber
\end{eqnarray}

\subsection{The Deviation vector}

Following Refs.~\cite{SLZ, NJfast,yan2021lamb}, the accuracy of our variational results can also be assessed by calculating the squared norm of the deviation vector (normalized by $\omega_0^2$), defined by
\begin{equation}
\begin{aligned}
\sigma^{2}(t) &= \frac{\|\,(i\partial_{t} - H) \,\ket{D_{2}^{M}(t)}\,\|^{2}}{\omega_{0}^{2}}\\
              &= \omega_{0}^{-2}\Bigl[\langle H^{2}\rangle \;-\;\bigl\langle \dot{D}_{2}^{M}(t)\bigl|\dot{D}_{2}^{M}(t)\bigr\rangle\Bigr].
\end{aligned}
\end{equation}
Detailed derivations are provided in Appendix~C. In General, the smaller the deviation \(\sigma^2(t)\), the more accurate the variational solutions are. In fact, \(\sigma^2(t) = 0\) if and only if the trial state is an exact solution to the Schr\''{o}dinger equation. Our earlier work has demonstrated that, provided $\sigma^2(t) < 10^{-2}$, the variational approach closely matches the hierarchy equations of motion \cite{yan2021lamb}. 
To illustrate this threshold, Fig.~2 shows \(\sigma^2(t)\) on a log–linear scale for a representative non-adiabatic benchmark (\(T=2\), \(\alpha/\omega_0=0.02\), \(\omega_{c}/\omega_0=1.5\)) at multiplicity \(M=18\).  The deviation remains below \(10^{-2}\) throughout the evolution, confirming that this trajectory faithfully tracks the exact dynamics, while in the non-convergent case(\(M=15\) the dashed curve crosses above \(10^{-2}\) around \(t\omega_0=3\), indicating a loss of precision.  Crucially, across all parameter sets and Ansatz multiplicities that yield convergent population dynamics, we find
\[
\max_{t}\,\sigma^2(t)\;<\;10^{-2},
\]
and all those convergent trajectories have passed this test in the main text.  Consequently, by monitoring \(\max_t\sigma^2(t)\), the validity of every variational result can be rigorously confirmed.

\begin{figure}[htbp]
  \centering
  \includegraphics[width=0.4\textwidth]{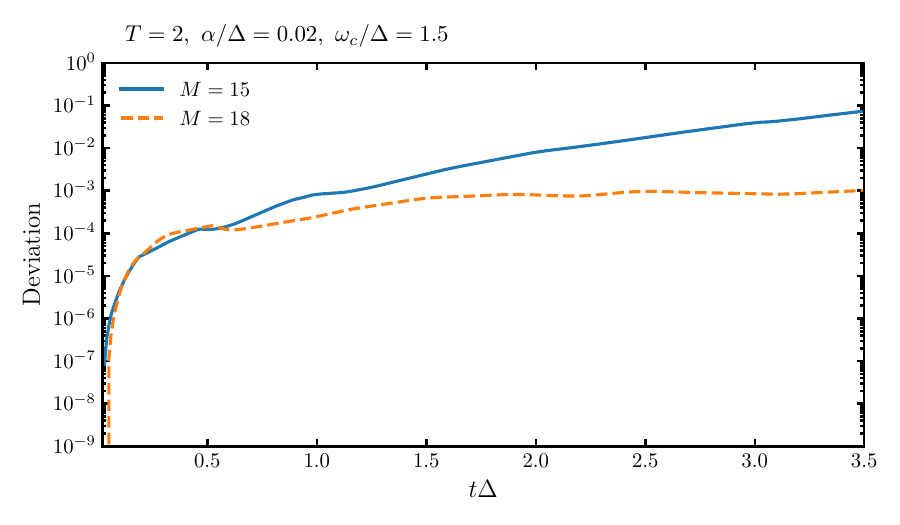}
  \caption{Time evolution of the deviation \(\sigma^2(t)\) on a log–linear scale for \(T=2\), \(\alpha/\omega_0=0.02\), \(\omega_{c}/\omega_0=1.5\).  The dashed horizontal line marks the convergence threshold \(\sigma^2=10^{-2}\).}
  \label{fig:deviation}
\end{figure}

\section{RESULTS AND DISCUSSION}
\subsection{Numerical aspects}
To benchmark our approach, we compared all three methods across every parameter regime of interest. In this context, we employ a temperature-corrected critical coupling formula
\begin{equation}
\alpha_{c}(T,\omega_{c}) = \alpha_{c}(0)\left(1 + \frac{k_{B}T}{\omega_{c}}\right)^{-1},
\end{equation}
which is derived assuming a well-defined cutoff frequency $\omega_{c}$ and based on low-order approximations.\cite{leggett1987dynamics} In the low-temperature limit ($k_{B}T \ll \omega_{c}$), this expression substantially reduces the effective coupling relative to its zero-temperature value. However, as the temperature approaches or exceeds $\omega_{c}$, the validity of this simplified formula becomes increasingly questionable. Although the mathematical limit $k_{B}T \gg \omega_{c}$ formally drives $\alpha_{c}(T,\omega_{c})$ toward zero, such an extreme high-temperature regime lies outside both the approximation's range of validity and our primary parameter space of interest. Consequently, we confine our analysis to scenarios where $k_{B}T$ does not significantly exceed $\omega_{c}$, ensuring that the temperature correction remains accurate and that higher-order thermal fluctuations or multi-body effects do not dominate the dynamics.

By applying this temperature correction within its appropriate domain, we can clearly identify weak-coupling, intermediate-coupling, and strong-coupling regimes at various temperatures. This classification not only highlights the role of temperature in governing the quantum dynamics of the system, but also serves as a unified framework for comparing numerical methods. With this scheme, we can more rigorously evaluate the performance of HEOM, QUAPI, and mDA at various temperatures, thereby assessing each method's accuracy and stability within its most suitable operational range. In particular, we define the intermediate coupling regime by
\[
0.1\,\alpha_{c}(0)\left(1+\frac{\omega_{c}}{k_{B}T}\right)^{-1} \le \alpha \le \alpha_{c}(0)\left(1+\frac{\omega_{c}}{k_{B}T}\right)^{-1}.
\]

Hence, as illustrated in Fig.~3,  we can visualize cuts through the full parameter space in the form of a two-dimensional ``phase diagram'' spanned by $\alpha$ and $\omega_c$, with the tunnelling parameter set to zero, $\Delta = 0$. Along the horizontal axis, shown on a logarithmic scale, lies the coupling strength (taken to be the same for both baths), while the vertical axis corresponds to the scaled cutoff frequency, $\omega_c$. The temperatures of the two baths are set to an average value $T$, with their difference fixed at $0.01\,T$.

\begin{figure}[htbp]
  \centering
  \includegraphics[width=0.5\textwidth]{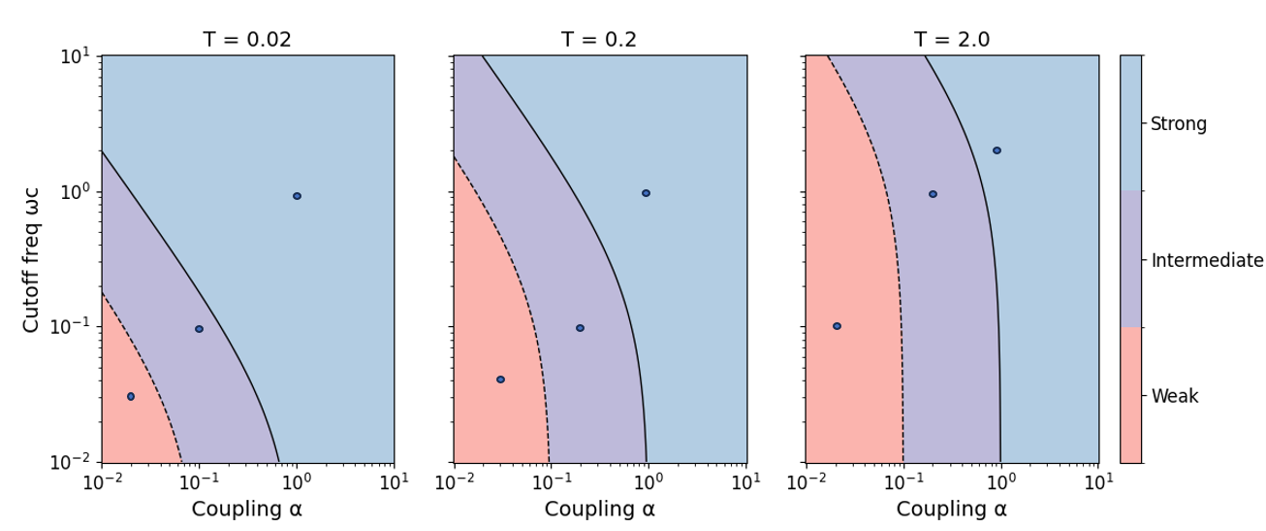}
  \caption{Parameter space of the model with zero tunneling parameter (\(\Delta = 0\)). The horizontal axis (in logarithmic scale) corresponds to the coupling strength, while the vertical axis (in linear scale) represents the cutoff frequency \(\omega_c\). Three temperature regimes are considered: (a) \(T = 0.02\) (very low temperature), (b) \(T = 0.2\) (low temperature), and (c) \(T = 2.0\) (high temperature). The marked points denote the specific parameter values used in our simulations.}
  \label{fig:my_label0}
\end{figure}

In the following computations, we set each simulation step to run for a fixed time step $\Delta t = 0.01/\omega_0$. Convergence is verified in two stages: first, we fix $N_b$ and gradually increase $M$ (the multiplicity) until further increases in $M$ produce negligible changes in the qubit's dynamics. Next, holding $M$ constant, we increase $N_b$ (the number of modes) until the qubit observables become insensitive to additional modes. The resulting variational data, referred to below as the mD2 results, have been thoroughly checked for convergence. For completeness, we also run multiple independent trajectories and compute the deviation vector to confirm the reliability of our methodology, shown in the lower panels of the figures. In addition to verifying convergence, we have also investigated the robustness of our mD2 approach in both the adiabatic ($\omega_c \ll \omega_0$) and non-adiabatic ($\omega_c \gtrsim \omega_0$) regimes. Specifically, we vary the cutoff frequency $\omega_c$ of the bath to compare the qubit's dynamics under slow- and fast-environment conditions. As expected, in the adiabatic regime, the environment behaves quasi-statically, and the qubit evolution shows only minor deviations even as $\omega_c$ decreases further. Conversely, in the non-adiabatic regime, where $\omega_c$ is comparable to or exceeds $\omega_0$, we observe more pronounced transient effects and faster energy exchanges between the qubit and its baths. These findings confirm that our methodology remains reliable over a wide range of bath speeds, consistently reproducing benchmark behaviors in both limits.

We compare our calculations with two established, numerically accurate methods: QUAPI and  HEOM~\cite{makarov1994path,makarov1995numerical,makarov1996long,tanimura1989,ishizaki2009}.
QUAPI discretizes the influence functional for propagation on the Keldysh contour, allowing control over the memory length $k_{\max}$ during propagation. If the system's memory is long, $k_{\max}$ must be large, causing convergence difficulties. HEOM, on the other hand, employs a Matsubara expansion of the bath and introduces a hierarchy of auxiliary density matrices. The truncation level $L$ and number of Matsubara terms $K$ are numerical parameters used to gain in convergence of the HEOM result. A standard and highly parallel HEOM implementation~\cite{struempfer2012} is known to be accurate for Debye spectral densities at high temperature but typically encounters challenges at lower temperatures or with non-Debye baths.

In this work, we use these methods primarily to gauge the accuracy of mDA, without attempting an exhaustive assessment of computational efficiency. Generally, on a typical laptop with our present (non-optimized) Python code, most mDA runs (which scale linearly with simulation time) can be completed within hours. Simulations in high-temperature regimes sometimes require a small cluster. Based on experience with related algorithms~\cite{dunnett2021matrix}, we estimate that an efficient or compiled implementation could further reduce the runtime by one to two orders of magnitude.

\subsection{High Temperature Regime}

We initiate our analysis by exploring the mDA approach in the high-temperature regime depicted in Fig.~3, with a focus on $k_B T = 2$. At \(k_{B}T = 2\), we explore the mDA across three characteristic coupling strengths—weak, intermediate, and strong —along the vertical slices in Fig.~3.  In each regime, we begin by comparing the population dynamics \(\langle\sigma_{z}(t)\rangle\) from mDA against benchmark QUAPI and HEOM results, proceed to determine the minimal multiplicity \(M\) required for stable trajectories, and finally interpret any residual discrepancies in terms of underlying physics. This unified narrative ensures a consistent comparison of accuracy, convergence, and mechanism across all three coupling regimes.

\subsubsection{Weak Coupling}

We begin by examining the weak-coupling regime ($\alpha = 0.02$). As shown in Fig.~4(b), both higher ($M = 18$) and lower ($M = 15$) multiplicities in the mD2 approach provide quantitatively accurate results, with the maximum deviation vector remaining on the order of $10^{-3}$ throughout the simulation. This demonstrates that, for moderate values of $M$, the mD2 method can capture the dynamics in this parameter regime with relatively low computational cost.

By contrast, when the cutoff frequency $\omega_c$ is large, the simulation requires more computational effort to maintain the same level of accuracy. This increased cost arises from the longer correlation times associated with low-$\omega_c$ environments, making the dynamics simulation a more demanding task. Nevertheless, all three methods---mD2, QUAPI, and HEOM---accurately capture the non-Markovian effects inherent in this regime, as evidenced by their close agreement.

We further note that for $M=15$ in non-adiabatic baths, the mD2 results remain reliable only up to $t \approx 3$. Beyond this point, the solution diverges from those obtained by QUAPI and HEOM, a discrepancy also evident in the deviation vector, which quickly rises above $10^{-1}$, as previously mentioned. The main reason is that, under fast-bath conditions, increasingly complex multi-phonon (multi-mode) processes arise over longer timescales, necessitating either a higher multiplicity $M$ or more refined computational resources to accurately track these numerous interference pathways. This drift is not due to the fact that, at longer times, the bath induces increasingly intricate multi-phonon (multi-mode) entanglement. Even so, in the high-temperature, weak-coupling regime, the mD2 approach still reproduces the HEOM benchmarks, and HEOM itself readily converges to exact solutions for the Debye spectral density under elevated temperatures. Meanwhile, QUAPI also achieves convergence in slow-bath conditions when a sufficiently large memory length is used, showing good agreement with both mD2 and HEOM.

\begin{figure}[htbp]
  \centering
  \includegraphics[width=0.5\textwidth]{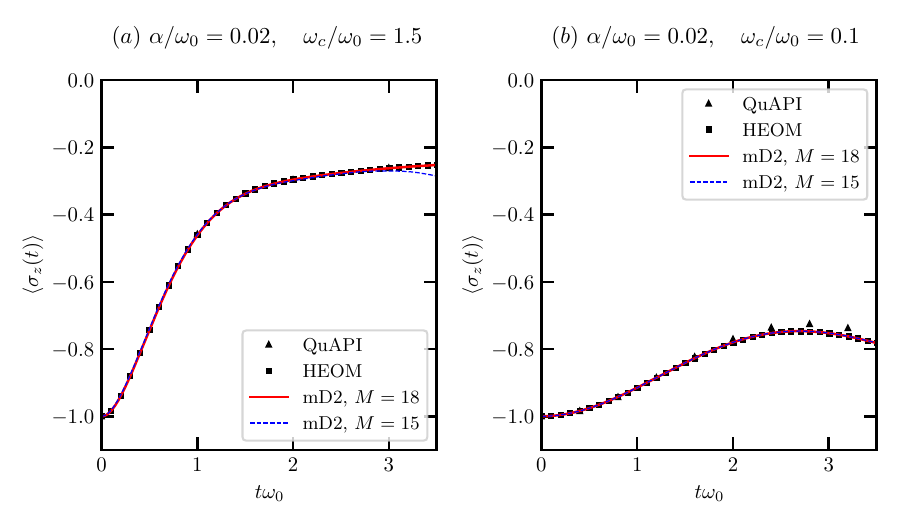}
  \caption{Time evolution of the nonequilibrium population difference $\sigma_z(t)$ for the weak coupling case ($\alpha=0.02$) at high temperature ($T=2.0$) with zero tunneling parameter ($\Delta=0$). The mD2 results (shown in red and orange) are presented for two scenarios: (a) a non-adiabatic (fast) bath with $\omega_c=1.5$, and (b) an adiabatic (slow) bath with $\omega_c=0.1$, with the mD2 multiplicity indicated by $M$. Triangles denote the QUAPI data (using $\Delta t=0.1/\omega_0$, $k_{\max}=7$ for (a) and $\Delta t=0.1/\omega_0$, $k_{\max}=9$ for (b)), and dashed lines mark the HEOM results obtained with $Nk=2$ and ${\rm max}_{\rm depth}=20$.}

  \label{fig:my_label1}
\end{figure}

\subsubsection{Strong Coupling}

We now move to the strong-coupling regime, where we anticipate that the mD2 method will be comparatively more challenging to converge. As shown in the upper panels of Fig.~5(a), the mD2 solution converges to a correct population profile upon increasing the multiplicity $M$, although reaching convergence appears to require at least $M=18$. While this result aligns with the expectation that convergence is difficult in this regime, it remains attainable with sufficient computational resources.

Contrary to the expectation that long-time memory effects in the adiabatic (slow) bath (\(\omega_c/\omega_0 = 0.5\)) would dictate convergence demands, our simulations demonstrate that it is in fact the non-adiabatic (fast) bath (\(\omega_c/\omega_0 = 2\)) that requires a larger \(M\). High cutoff frequencies drive rapid multi-phonon excitations and foster intricate mode-mode interference—dynamics that a limited mD2 manifold cannot span. Only by increasing \(M\) to 20 (or higher) does the Ansatz recover precise agreement with QUAPI and HEOM. In contrast, the slow bath’s low-frequency spectrum produces a quasi-static polarization that is already captured at modest multiplicities, rendering further increases in \(M\) largely superfluous.

These convergence difficulties arise from an enlarged effective Hilbert space, stronger qubit–phonon interactions, and a more intricate energy landscape, each of which amplifies numerical instabilities in the evolution of variational parameters. Despite these hurdles, mD2 can still achieve perfect agreement with HEOM benchmarks—maintaining deviation vectors below \(10^{-2}\)—whereas QUAPI fails to converge, as evidenced by the discrepancies in Fig.~5.

The rapid oscillatory “ripples” with amplitudes of \(10^{-2}\) superimposed on the principal coherent-decay envelope originate from the discrete bath-mode representation of the continuous Drude–Lorentz spectral density. Each discretized mode \(\omega_k\) contributes a term proportional to \(\cos(\omega_k t)\) to the system’s autocorrelation, yielding high-frequency oscillations at short times (\(t \lesssim 1/\omega_c\)). Because mode density decreases near the cutoff, these contributions do not fully cancel, producing a visible beating pattern rather than a smooth decay. A Fourier transform of \(\langle\sigma_z(t)\rangle\) confirms that the spectral peaks align precisely with \(\omega_k\) and \(\omega_c\), demonstrating that these features are inherent to the discretized bath rather than numerical artifacts. Moreover, the amplitude of these high-frequency components scales with the coupling strength \(\eta\) and is modulated by the thermal factor \(\coth(\beta\omega_k/2)\), making them most pronounced under strong coupling or elevated temperature. Hence, these characteristic oscillations are only observable under the fast‐bath, strong‐coupling conditions as in Fig.~5.

\begin{figure}[htbp]
  \centering
  \includegraphics[width=0.5\textwidth]{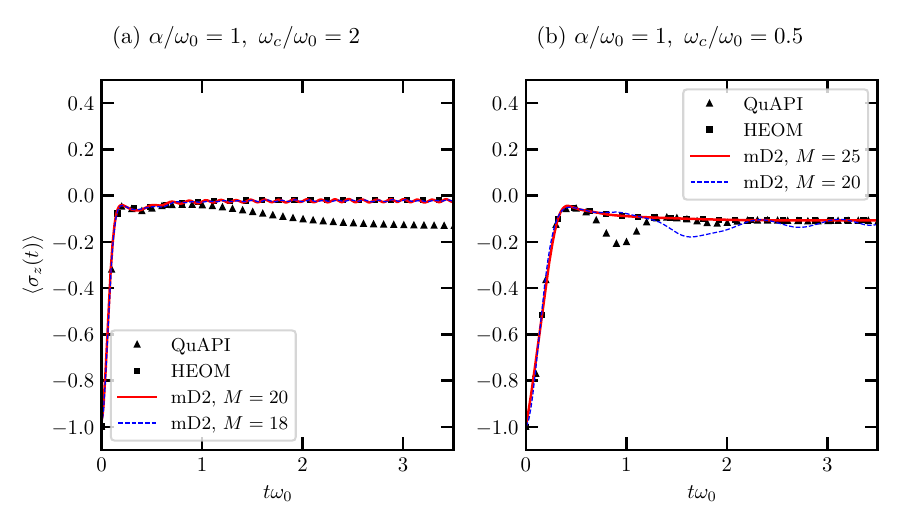}
  \caption{Time evolution of the nonequilibrium population difference $\sigma_z(t)$ for the strong coupling case ($\alpha=1.0$) at high temperature ($T=2.0$) with zero tunneling parameter ($\Delta=0$). The mD2 results (shown in red and orange) are presented for two scenarios: (a) a non-adiabatic (fast) bath with $\omega_c=2$, and (b) an adiabatic (slow) bath with $\omega_c=0.5$, with the mD2 multiplicity indicated by $M$. Triangles denote the QUAPI data (using $\Delta t=0.1/\omega_0$, $k_{\max}=8$ for (a) and $\Delta t=0.3/\omega_0$, $k_{\max}=11$ for (b)), and dashed lines mark the HEOM results obtained with $Nk=2$ and ${\rm max}_{\rm depth}=20$.}
  \label{fig:my_label2}
\end{figure}

\subsubsection{Intermediate Coupling}

\begin{figure}[htbp]
  \centering
  \includegraphics[width=0.5\textwidth]{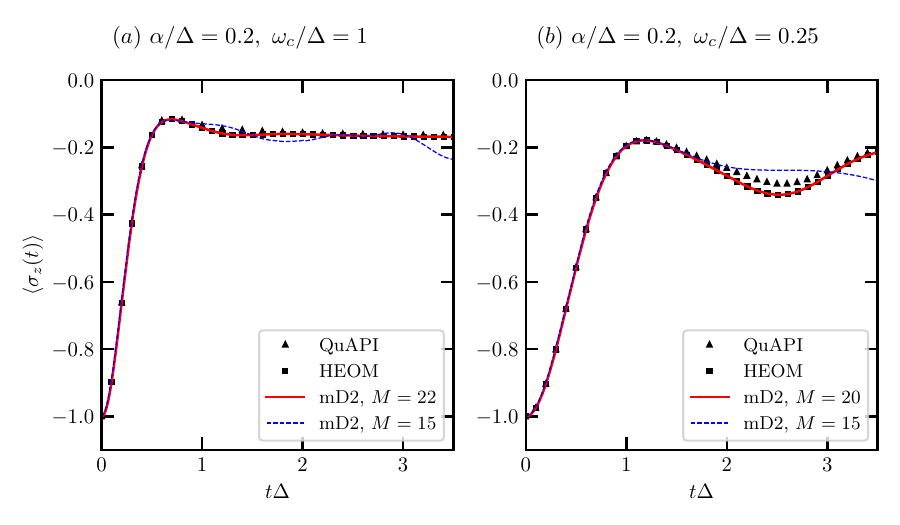}
  \caption{Time evolution of the nonequilibrium population difference $\sigma_z(t)$ for the intermediate coupling case ($\alpha=0.2$) at high temperature ($T=2.0$) with zero tunneling parameter ($\Delta=0$). The mD2 results (shown in red and orange) are presented for two scenarios: (a) a non-adiabatic (fast) bath with $\omega_c=1$, and (b) an adiabatic (slow) bath with $\omega_c=0.25$, with the mD2 multiplicity indicated by $M$. Triangles denote the QUAPI data (using $\Delta t=0.1/\omega_0$, $k_{\max}=6$ for (a) and $\Delta t=0.2/\omega_0$, $k_{\max}=10$ for (b)), and dashed lines mark the HEOM results obtained with $Nk=2$ and ${\rm max}_{\rm depth}=20$.}
  \label{fig:my_label3}
\end{figure}

Finally, we consider the intermediate-coupling regime with coupling strength $\alpha = 0.2$. As expected, the required multiplicity $M$ falls between the weak- and strong-coupling cases. For the fast bath ($\omega_c/\omega_0 = 1$), $M=15$ already reproduces the exact dynamics (QUAPI/HEOM) accurately up to $t\omega_0 \approx 3.5$, as shown in panel (a) of Fig.~\ref{fig:my_label3}. In contrast, for the slow bath ($\omega_c/\omega_0 = 0.25$), the long-lived correlations of low-frequency modes induce larger deviations even at $M=18$; these arise from multi-phonon interference that accumulates over the extended bath memory time, as depicted in panel (b) of Fig.~\ref{fig:my_label3}. However, systematic increases of $M$ rapidly restore quantitative agreement. Notably, in this intermediate coupling regime, QUAPI and HEOM trajectories overlap without requiring an extremely fine time discretization or an excessively long memory kernel—because the bath correlation time lies within the discretization and memory cutoff already employed.

\subsection{Low Temperature Regime}

Turning to the low-temperature portion of the phase diagram, we specifically set $k_B T = 0.2$ and focus on vertical cuts at both intermediate and strong coupling, corresponding to Fig.~3(b).

\subsubsection{Intermediate Coupling}

In the intermediate‐coupling regime \(\alpha/\omega_0=0.2\), mD2 demonstrates markedly faster convergence at low temperatures due to the suppression of thermally activated high‐frequency bath excitations and reduced multi‐phonon interference: as shown in Fig.~7, a modest multiplicity \(M=10\) already suffices to match QUAPI and HEOM dynamics for both \((\alpha,\omega_c)=(0.2,1.5)\) and \((0.2,0.1)\), reproducing the initial rapid rise, overdamped relaxation, and underdamped oscillations to within \(10^{-3}\) accuracy.  In this low‐\(T\) regime, the thermal prefactor \(\coth(\beta\omega/2)\) approaches unity for \(\omega\gtrsim k_BT\), so the effective bath memory kernel is dominated by its zero‐temperature form and its short‐time high‐frequency tail decays smoothly, minimizing the need for high \(M\) to resolve complex interference pathways.  By contrast, in the high‐temperature scenarios of Sec.~III\thinspace B, \(\coth(\beta\omega/2)\gg1\) for low‐to‐mid frequencies, amplifying a broad spectrum of modes and necessitating larger multiplicity (or computational effort) to achieve comparable error levels.  This distinction is quantitatively confirmed by the deviation vector \(\|\sigma(t)^2\|\), which remains on the order of \(10^{-3}\) at low \(T\) but grows more rapidly when \(k_BT\) increases, underscoring the enhanced “exactness” of the variational solution under colder bath conditions.

\begin{figure}[htbp]
  \centering
  \includegraphics[width=0.5\textwidth]{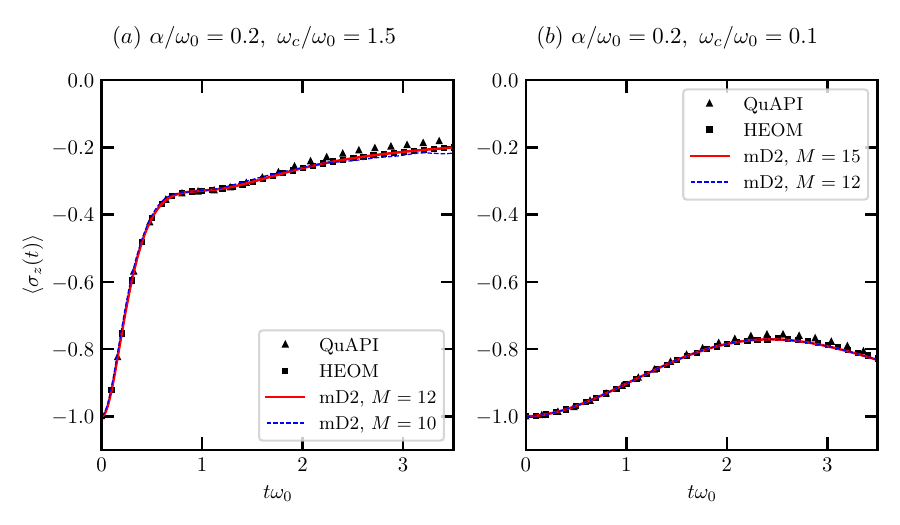}
  \caption{Time evolution of the nonequilibrium population difference $\sigma_z(t)$  for the intermediate coupling case ($\alpha=0.2$) at low temperature ($T=0.2$) with zero tunneling parameter ($\Delta=0$). The mD2 results (shown in red and orange) are presented for two scenarios: (a) a non-adiabatic (fast) bath with $\omega_c=1.5$, and (b) an adiabatic (slow) bath with $\omega_c=0.1$, with the mD2 multiplicity indicated by $M$. Triangles denote the QUAPI data (using $\Delta t=0.1/\omega_0$, $k_{\max}=6$ for (a) and $\Delta t=0.1/\omega_0$, $k_{\max}=10$ for (b)), and dashed lines mark the HEOM results obtained with $Nk=2$ and ${\rm max}_{\rm depth}=20$.}
  \label{fig:my_label4}
\end{figure}

\subsubsection{Strong Coupling}

\begin{figure}[htbp]
  \centering
  \includegraphics[width=0.5\textwidth]{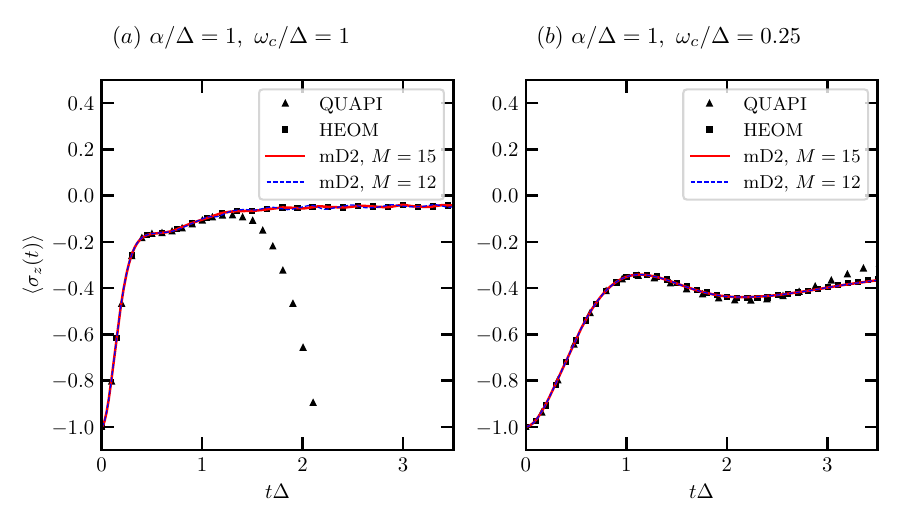}
  \caption{Time evolution of the nonequilibrium population difference $\sigma_z(t)$ for the strong coupling case ($\alpha=1$) at low temperature ($T=0.2$) with zero tunneling parameter ($\Delta=0$). The mD2 results (shown in red and orange) are presented for two scenarios: (a) a non-adiabatic (fast) bath with $\omega_c=1$, and (b) an adiabatic (slow) bath with $\omega_c=0.25$, with the mD2 multiplicity indicated by $M$. Triangles denote the QUAPI data (using $\Delta t=0.2/\omega_0$, $k_{\max}=10$ for (a) and $\Delta t=0.3/\omega_0$, $k_{\max}=11$ for (b)), and dashed lines mark the HEOM results obtained with $Nk=3$ and ${\rm max}_{\rm depth}=20$.}
  \label{fig:my_label5}
\end{figure}

In the strong‐coupling regime $\alpha/\omega_0=1$, the mD2 method continues to yield quantitatively accurate results even for large cutoff frequencies, but at the cost of increased variational complexity: for $\omega_c/\omega_0=1$, achieving full convergence requires a multiplicity as high as $M=15$, which—using our current prototype implementation—translates into on the order of ten CPU‐hours per trajectory. This computational expense reflects the need to resolve an extensive network of multi‐phonon interference pathways that develop at strong spin–bath coupling. By comparison, QUAPI systematically overestimates the long‐time decay of $\langle\sigma_z(t)\rangle$ in both adiabatic and non‐adiabatic baths, a symptom of its finite memory cutoff $k_{\max}$. Extending QUAPI’s long‐time accuracy would necessitate increasing $k_{\max}$ substantially, but since the path‐integral tensor dimension scales roughly as $d^{2k_{\max}}$ (with $d$ the system Hilbert‐space dimension), even modest increases in $k_{\max}$ produce an exponential rise in memory footprint, rendering convergence impractical for $k_{\max}\gtrsim50$. Thus, while both methods face challenges in the extended‐time, strong‐coupling limit, mD2 affords a more controlled trade‐off between accuracy and runtime via the single parameter $M$, whereas QUAPI’s memory demands grow prohibitively large, as illustrated in Fig.~\ref{fig:my_label5}.

\subsection{The Very Low Temperature Limit}

We now turn to the very low temperature limit, taking $k_B T = 0.02$ as specified by the phase diagram in Fig.~2(c). In this regime, a standard HEOM implementation becomes computationally expensive, since decreasing the bath temperature generally demands more Matsubara terms to accurately encode the bath density matrix, along with additional hierarchical levels to ensure convergence at long times. Indeed, we found that our current HEOM setup becomes prohibitively costly at such low temperatures, although more recent techniques may mitigate these issues in certain scenarios~\cite{hu2010,hu2011,yan2014,tang2015,ye2016}.

Fig.9 indicates that mD2 retains an advantage over both HEOM and Quapi at low temperatures across all coupling strengths, owing to the fact that its computational expense does not grow significantly with decreasing temperature. For a rapid bath ($\omega_c = 1.5$), as illustrated in Fig.~9(a), mD2 maintains accurate population dynamics while also showing higher efficiency than the Quapi method. Conversely, in the intermediate cutoff-frequency regime ($\omega_c = 0.1$), mD2 yields population evolutions that align closely with QUAPI, consistent with our earlier estimates regarding the method's convergence properties.
\begin{figure}[htbp]
 \centering
 \includegraphics[width=0.5\textwidth]{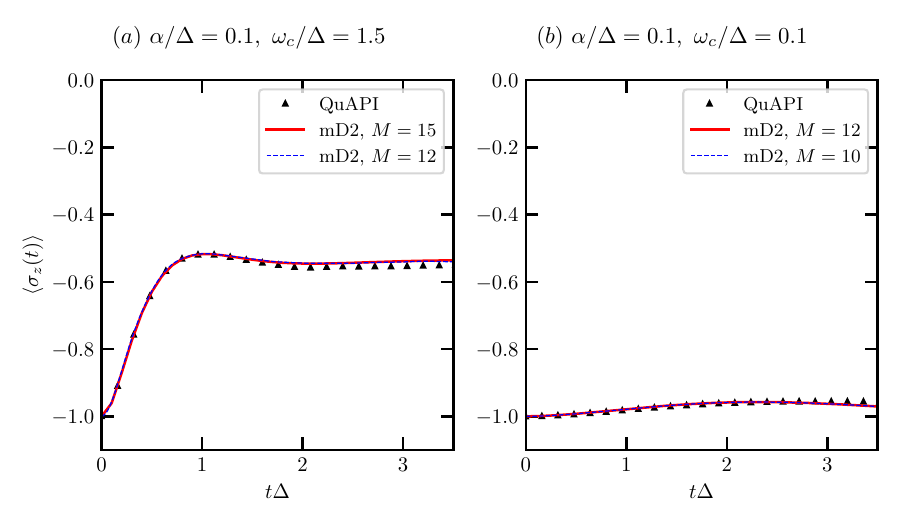}
 \caption{Time evolution of the nonequilibrium population difference $\sigma_z(t)$ for the intermediate coupling case ($\alpha=0.1$) at very low temperature ($T=0.02$) with zero tunneling parameter ($\Delta=0$). The mD2 results (shown in red and orange) are presented for two scenarios: (a) a non-adiabatic (fast) bath with $\omega_c=1.5$, and (b) an adiabatic (slow) bath with $\omega_c=0.1$, with the mD2 multiplicity indicated by $M$. Triangles denote the QUAPI data (using $\Delta t=0.2/\omega_0$, $k_{\max}=10$ for (a) and $\Delta t=0.1/\omega_0$, $k_{\max}=10$ for (b)), and the HEOM results are no longer avaliable in this regime.}
 \label{fig:my_label6}
\end{figure}

\subsection{Scaling behavior}

Assessing the computational scaling of various numerical methods is essential for determining their feasibility in simulating large systems or long-time dynamics, as well as for properly interpreting the results discussed earlier. In this subsection, we examine the scaling behavior of the three aforementioned methods.

QUAPI employs a discretized path-integral approach to capture non-Markovian effects by retaining information from the last \( L \) time steps. For a system with a Hilbert space of dimension \( d \), the density matrix is \( d \times d \). However, since QUAPI keeps track of the previous \( L \) time steps, the effective state space becomes a tensor product of \( L \) density matrices, leading to a scaling of \( \mathcal{O}\big(d^{2L}\big) \) \cite{bose2023quantumdynamics}. This exponential dependence on the memory length \( L \) becomes a major bottleneck, especially for systems with long bath correlation times or strong system-bath coupling. 

Recent improvements to QUAPI, including the integration with tensor network methods like the Time-Evolved Matrix Product Operators (TEMPO) method \cite{strathearn2018efficient} and the Multi-Site Tensor Network Path Integral (MS-TNPI) method \cite{bose2022multisite}, have significantly reduced the computational complexity. These methods scale polynomially with system size and memory length, making them more efficient for large systems and long-time simulations. Despite these advancements, in this manuscript, we choose to use the unoptimized form of QUAPI for consistency in comparison with other methods, as it allows for a direct and fair evaluation of the methods in their basic, unoptimized versions.

HEOM, on the other hand, reformulates the environmental influence in terms of a hierarchy of auxiliary density operators (ADOs). If the bath correlation function is decomposed into $K$ Matsubara terms and the hierarchy is truncated at level $L$, the total number of ADOs is estimated as \cite{kreisbeck2011high}
\[
N_{\text{ADO}} \approx \binom{K+L}{L} = \frac{(K+L)!}{K!\,L!}\,.
\]
Since each ADO is a $d\times d$ matrix, the overall computational cost for storing and propagating these operators scales roughly as
\[
\mathcal{O}\Bigl(d^2 \times \frac{(K+L)!}{K!\,L!}\Bigr).
\]
In practice, particularly at low temperatures or under strong coupling conditions, both $K$ and $L$ may need to be large, leading to a rapid, often nearly exponential growth in computational cost.

In contrast, the mDA method represents the system's wave function as a superposition of $M$ states. For each state, the variational parameters include the system amplitudes as well as the displacement parameters for each of the $N$ bath modes. The total number of variational parameters is approximately
\[
N_{\text{var}} \times M \times N.
\]
During time propagation (typically via a Runge-Kutta scheme), one must compute the equations of motion for all these parameters. Additionally, the evaluation of overlaps (e.g., Debye-Waller factors) between different multiplets introduces a cost that scales as $\mathcal{O}(M^2)$. Thus, the computational effort per time step scales as
\[
\mathcal{O}\left(M^2 \times N\right).
\]
This polynomial scaling in both $M$ and $N$ is considerably more favorable than the exponential scaling of QUAPI and the combinatorial scaling of HEOM.

For example, in the high-temperature regime, the mDA method may require \( M = 20 \) and \( N = 60 \) for convergence, leading to a computational complexity of approximately \( \mathcal{O}(19,440) \), meaning it requires around 19,440 operations. This is significantly higher than HEOM, with a complexity of about \( \mathcal{O}(924) \) (924 operations), and QUAPI, with \( \mathcal{O}(4,096) \) (4,096 operations). However, at lower temperatures, the required \( M \) can be reduced, resulting in a complexity of about \( \mathcal{O}(8,640) \) (8,640 operations), which is closer to HEOM (\( \mathcal{O}(7,084) \)) and much more efficient than QUAPI (\( \mathcal{O}(262,144) \)). In very low-temperature regimes, mDA becomes more favorable due to the reduced complexity. Additionally, using more efficient discretization techniques can further decrease the computational cost, improving mDA's performance for larger or more complex systems.

\section{Conclusion}

In this study, we have systematically benchmarked the mD2 approach for modeling open quantum systems at finite temperatures. By integrating mD2 with TFD, we investigated a two-reservoir model where each reservoir has a distinct temperature. Through detailed comparisons with established numerical methods such as HEOM and QUAPI, we demonstrated that mDA achieves accurate results across a broad range of coupling strengths, spectral densities, and temperature regimes. Notably, mDA retains its accuracy and stability even in challenging parameter spaces where conventional techniques struggle with convergence or computational overhead.

Our findings underscore the adaptability of mDA as a robust framework for simulating quantum thermodynamics and non-equilibrium quantum systems. While QUAPI and HEOM are undoubtedly well-established, effective tools in their respective domains of applicabilities, mDA offers unique advantages in handling strong system-bath coupling and capturing long-time dynamics with manageable computational cost. Future studies may explore further optimizations of the mDA method, including enhanced numerical implementations and extensions to multi-level quantum systems and non-Hermitian dynamics, and applications of mDA to complex physical systems, such as quantum heat engines and organic photovoltaic cells.


\section*{Acknowledgments}

The authors thank Maxim Gelin, Yiying Yan, and Xuanhua Zhang for useful discussion. Support of the Singapore Ministry of Education Academic Research Fund (Grant No.~RG2/24) is gratefully acknowledged. Lu Wang would like to thank National Natural Science Foundation of China for partial support (Grant No. 22163005).
F. Z. acknowledges support from "Yongjiang Talents Program” under Grant No. 2023A-158-G funded by the Ningbo Science and Technology Bureau.

\hspace{0.5cm}

\section*{Conflict of Interest}
The authors have no conflicts to disclose.

\section*{Data Availability}
The data that support the findings of this study are available from the corresponding author upon reasonable request.

\appendix
\onecolumngrid
\section{Discretization of the Bosonic Bath via Interpolative Decomposition}\label{Appendix A}
In this appendix, we follow the Ref.\,\cite{takahashi2024discretization} and describe in detail the procedure used to discretize the continuous spectral density of the bosonic bath. Our starting point is the bath correlation function (BCF)
\[
C(t)=\int_{-\Omega}^{\Omega} S_{\beta}(\omega)e^{-i\omega t}\,d\omega,
\]
where the quantum noise spectral density is defined as
\[
S_\beta(\omega) \equiv \frac{1}{2\pi}\,J(\omega) \left[\coth\left(\frac{\beta\omega}{2}\right) + 1\right]\,.
\]
and we wish to approximate this BCF by a discrete sum of the form
\[
C(t)\approx \sum_{k=1}^{M} z_k\,S_{\beta}(\omega_k)e^{-i\omega t}\,.
\]
To achieve this, we first discretize the frequency domain \( [-\Omega,\Omega] \) into \( n \) equispaced points \(\{\omega_j\}_{j=1}^{n}\) and the time domain \([0,T]\) into \( m \) points \(\{t_i\}_{i=1}^{m}\). The BCF at the discrete times is then approximated using a quadrature rule:
\[
C(t_i)\approx \sum_{j=1}^{n} f(t_i,\omega_j)\,w_j,\quad\text{with}\quad f(t,\omega)=S_{\beta}(\omega)e^{-i\omega t}\,,
\]
where \(\{w_j\}\) are the quadrature weights.

Next, we form the \( m\times n \) complex matrix \( F \) with elements
\[
F_{ij}=S_{\beta}(\omega_j)e^{-i\omega_j t_i}\,.
\]
To facilitate a low-rank approximation, we construct the real matrix
\[
F_{\text{real}}=\begin{pmatrix}\Re\,F\\[2mm]\Im\,F\end{pmatrix}\in\mathbb{R}^{2m\times n}\,.
\]
An interpolative decomposition (ID) is then performed on \( F_{\text{real}} \) so that
\[
F_{\text{real}}=B\,P+E\,,
\]
where \( B\in\mathbb{R}^{2m\times r} \) is a matrix formed by selecting \( r \) representative columns from \( F_{\text{real}} \), \( P\in\mathbb{R}^{r\times n} \) is the corresponding interpolation matrix, and the error matrix \( E\in\mathbb{R}^{2m\times n} \) satisfies \(\|E\|\le \epsilon\) for a chosen tolerance \(\epsilon\). The selected column indices define a subset \(\widetilde{\Omega}=\{\omega_k\}_{k=1}^{r}\subset\{\omega_j\}\) that captures the dominant frequency components.

Substituting the ID approximation into the quadrature expression yields an approximate representation of the BCF:
\[
C(t_i)\approx \sum_{k=1}^{r} S_{\beta}(\omega_k)e^{-i\omega_k t_i}\,z_k\,,
\]
with the effective weights defined by
\[
z_k=\sum_{j=1}^{n}P_{kj}\,w_j\,.
\]
In practice, the coefficients \( \{z_k\} \) are obtained by constructing the data vector
\[
c=\begin{pmatrix}\Re\,C(t_1)\\[2mm]\vdots\\[2mm]\Re\,C(t_m)\\[2mm]\Im\,C(t_1)\\[2mm]\vdots\\[2mm]\Im\,C(t_m)\end{pmatrix}\in\mathbb{R}^{2m}
\]
and solving the overdetermined system
\[
\min_{z\ge 0}\,\|c-B\,z\|_2\,,
\]
using a nonnegative least squares (NNLS) algorithm. The nonnegativity of \(z\) ensures that the derived effective system-bath coupling parameters
\[
g_k(\beta)=\sqrt{z_k\,S_{\beta}(\omega_k)}
\]
are real, thereby preserving the Hermiticity of the discrete bath Hamiltonian.

In our simulations the frequency interval \([-\Omega,\Omega]\) is chosen sufficiently wide so that contributions outside this interval are negligible. The discrete frequencies \(\{\omega_k\}\) and the corresponding coupling strengths \(\{g_k(\beta)\}\) obtained through this procedure are then used in the variational treatment of the open quantum system dynamics.

\begin{figure}[htbp]
  \centering
  \includegraphics[width=0.8\linewidth]{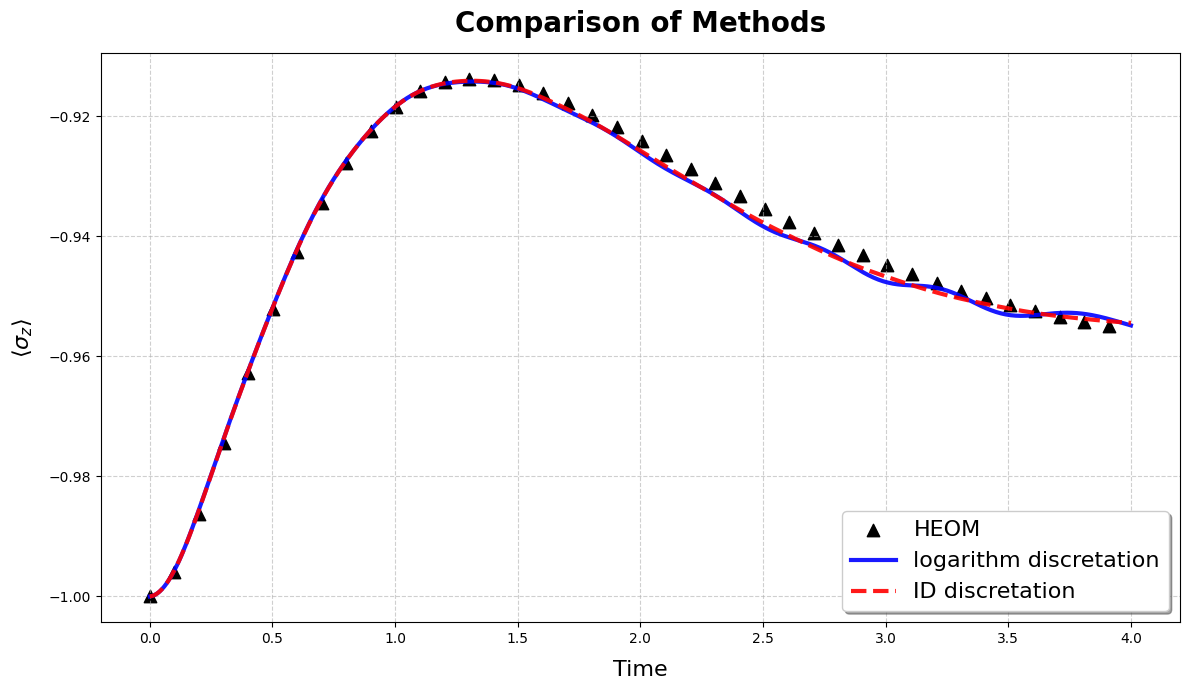}
  \caption{Comparison of the population inversion $\langle\sigma_z\rangle$ obtained from mD2 calculations using logarithmic discretization (blue line, 60 modes) and ID discretization (red dashed line, 28 modes), alongside the HEOM benchmark (black triangles). The ID approach not only reproduces the long-time dynamics more accurately but also avoids the spurious oscillations seen with the logarithmic discretization.}
  \label{fig:ID_comparison}

\end{figure}

Furthermore, we compare the performance of our mD2 approach when employing two different bath-discretization schemes: the logarithmic discretization and the interpolative decomposition (ID) method. The Fig.~\ref{fig:ID_comparison} shows that the ID approach yields more accurate results, especially at longer times, while requiring fewer discrete bath modes (28 modes vs 60 modes) than the logarithmic method. Moreover, the logarithmic method tends to introduce some unphysical oscillation at longer time scales.

\clearpage
\section{The Equation of Motions for the Multi-D$_2$ Ansatz}\label{Appendix B}
\begin{equation}
    \begin{aligned}
        0=&  i \hbar\sum_{i, j}^{M} \sum_{k=1}^{N} \dot{A_{i}} S_{i,j} - i \hbar\sum_{i, j}^{M} \sum_{k=1}^{N}  A_{i}\left(\frac{\dot{f}_{j k}^{*} f_{j k} + f_{j k}^{*} \dot{f}_{j k} - \dot{f}_{i k} f_{i k}^{*} - f_{i k} \dot{f}_{i k}^{*}}{2} + f_{j k}^{*} \dot{f}_{i k} - f_{i k} \dot{f}_{j k}^{*}\right) S_{i,j}\\
        &+ i \hbar\sum_{i, j}^{M} \sum_{k=1}^{N}  A_{i}\left(\frac{\dot{\tilde{f}}_{j k}^{*} \tilde{f}_{j k} + \tilde{f}_{j k}^{*} \dot{\tilde{f}}_{j k} - \dot{\tilde{f}}_{i k} \tilde{f}_{i k}^{*} - \tilde{f}_{i k} \dot{\tilde{f}}_{i k}^{*}}{2} + \tilde{f}_{j k}^{*} \dot{\tilde{f}}_{i k} - \tilde{f}_{i k} \dot{\tilde{f}}_{j k}^{*}\right) S_{i,j}\\
        &-\sum_{i, j}^{M}  \frac{\omega_{0}}{2}A_{i}S_{i,j} - \omega_{k}\sum_{i, j}^{M} \sum_{k=1}^{N}A_{i}(f_{i,k}f_{j,k}^{*}+\tilde{f}_{i,k}\tilde{f}_{j,k}^{*})S_{i,j} - \lambda_{k}^{'}\sum_{i, j}^{M} \sum_{k=1}^{N}B_{i}\Bigl[(f_{j,k}+f_{j,k}^{*})+(\tilde{f}_{j,k}+\tilde{f}_{j,k}^{*})\Bigr]S_{i,j}\,.
    \end{aligned}
\end{equation}
\begin{equation}
    \begin{aligned}
        0=&  i \hbar\sum_{i, j}^{M} \sum_{k=1}^{N} \dot{B_{i}} S_{i,j} - i \hbar\sum_{i, j}^{M} \sum_{k=1}^{N}  B_{i}\left(\frac{\dot{f}_{j k}^{*} f_{j k} + f_{j k}^{*} \dot{f}_{j k} - \dot{f}_{i k} f_{i k}^{*} - f_{i k} \dot{f}_{i k}^{*}}{2} + f_{j k}^{*} \dot{f}_{i k} - f_{i k} \dot{f}_{j k}^{*}\right) S_{i,j}\\
        &+ i \hbar\sum_{i, j}^{M} \sum_{k=1}^{N}  B_{i}\left(\frac{\dot{\tilde{f}}_{j k}^{*} \tilde{f}_{j k} + \tilde{f}_{j k}^{*} \dot{\tilde{f}}_{j k} - \dot{\tilde{f}}_{i k} \tilde{f}_{i k}^{*} - \tilde{f}_{i k} \dot{\tilde{f}}_{i k}^{*}}{2} + \tilde{f}_{j k}^{*} \dot{\tilde{f}}_{i k} - \tilde{f}_{i k} \dot{\tilde{f}}_{j k}^{*}\right) S_{i,j}\\
        &- \sum_{i, j}^{M}  \frac{\omega_{0}}{2}B_{i}S_{i,j} + \omega_{k}\sum_{i, j}^{M} \sum_{k=1}^{N}B_{i}(f_{i,k}f_{j,k}^{*}+\tilde{f}_{i,k}\tilde{f}_{j,k}^{*})S_{i,j} - \lambda_{k}^{'}A_{i}\sum_{i, j}^{M} \sum_{k=1}^{N}\Bigl[(f_{j,k}+f_{j,k}^{*})+(\tilde{f}_{j,k}+\tilde{f}_{j,k}^{*})\Bigr]S_{i,j}\,.
    \end{aligned}
\end{equation}
\begin{equation}
    \begin{aligned}
        0=&  i \hbar \sum_{i, j}^{M} \sum_{k=1}^{N}\Bigl(A_{j}^{*} \dot{A}_{i}+B_{j}^{*} \dot{B}_{i}\Bigr)
        \Bigl(-f_{i,k}+\frac{1}{2}f_{j,k}\Bigr)S_{i,j} - i\hbar\sum_{i, j}^{M} \sum_{k=1}^{N}\Bigl(A_{j}^{*} A_{i}+B_{j}^{*} B_{i}\Bigr)\dot{f}_{i,k}S_{i,j}\\[2mm]
        &+ i\hbar\sum_{i, j}^{M} \sum_{k=1}^{N}\Bigl(A_{j}^{*} A_{i}+B_{j}^{*} B_{i}\Bigr)
        \Bigl(\dot{f}_{i,k} f_{j,k}^{*} - \frac{\dot{f}_{i,k} f_{i,k}^{*} + f_{i,k} \dot{f}_{i,k}^{*}}{2} + \dot{\tilde{f}}_{i,k} \tilde{f}_{j,k}^{*} - \frac{\dot{\tilde{f}}_{i,k} \tilde{f}_{i,k}^{*} + \tilde{f}_{i,k} \dot{\tilde{f}}_{i,k}^{*}}{2}\Bigr)\\[2mm]
        &- \sum_{i, j}^{M} \sum_{k=1}^{N}\frac{\omega_{0}}{2}  \Bigl(A_j^{*} A_i - B_j^{*} B_i\Bigr)  \Bigl(f_{i,k}-\frac{1}{2}f_{j,k}\Bigr)S_{i,j}\\[2mm]
        &- \sum_{i, j}^{M} \sum_{k=1}^{N} \omega_{k}\Bigl(A_j^{*} A_i + B_j^{*} B_i\Bigr)f_{i,k}S_{i,j}\\[2mm]
        &- \sum_{i, j}^{M} \sum_{k=1}^{N} \omega_{k}\Bigl(A_j^{*} A_i + B_j^{*} B_i\Bigr)
        \Bigl(f_{i,k}f_{j,k}^{*}+\tilde{f}_{i,k}\tilde{f}_{j,k}^{*}\Bigr)
        \Bigl(f_{i,k}-\frac{1}{2}f_{j,k}\Bigr)S_{i,j}\\[2mm]
        &- \sum_{i, j}^{M} \sum_{k=1}^{N} \lambda_{k}^{'} \Bigl(B_j^{*} A_i + A_j^{*} B_i\Bigr)\Bigl[(f_{j,k}+f_{j,k}^{*})+(\tilde{f}_{j,k}+\tilde{f}_{j,k}^{*})\Bigr]
        \Bigl(f_{i,k}-\frac{1}{2}f_{j,k}\Bigr)S_{i,j}\\[2mm]
        &- \sum_{i, j}^{M} \sum_{k=1}^{N} \lambda_{k}^{'} \Bigl(B_j^{*} A_i + A_j^{*} B_i\Bigr)S_{i,j}\,.
    \end{aligned}
\end{equation}
\begin{equation}
    \begin{aligned}
        0=&  i \hbar \sum_{i, j}^{M} \sum_{k=1}^{N}\Bigl(A_{j}^{*} \dot{A}_{i}+B_{j}^{*} \dot{B}_{i}\Bigr)
        \Bigl(-\tilde{f}_{i,k}+\frac{1}{2}\tilde{f}_{j,k}\Bigr)S_{i,j} - i\hbar\sum_{i, j}^{M} \sum_{k=1}^{N}\Bigl(A_{j}^{*} A_{i}+B_{j}^{*} B_{i}\Bigr)\dot{\tilde{f}}_{i,k}S_{i,j}\\[2mm]
        &+ i\hbar\sum_{i, j}^{M} \sum_{k=1}^{N}\Bigl(A_{j}^{*} A_{i}+B_{j}^{*} B_{i}\Bigr)
        \Bigl(\dot{f}_{i,k} f_{j,k}^{*} - \frac{\dot{f}_{i,k} f_{i,k}^{*} + f_{i,k} \dot{f}_{i,k}^{*}}{2} + \dot{\tilde{f}}_{i,k} \tilde{f}_{j,k}^{*} - \frac{\dot{\tilde{f}}_{i,k} \tilde{f}_{i,k}^{*} + \tilde{f}_{i,k} \dot{\tilde{f}}_{i,k}^{*}}{2}\Bigr)\\[2mm]
        &- \sum_{i, j}^{M} \sum_{k=1}^{N}\frac{\omega_{0}}{2} \Bigl(A_j^{*} A_i - B_j^{*} B_i\Bigr)
        \Bigl(\tilde{f}_{i,k}-\frac{1}{2}\tilde{f}_{j,k}\Bigr)S_{i,j}\\[2mm]
        &- \sum_{i, j}^{M} \sum_{k=1}^{N} \omega_{k}\Bigl(A_j^{*} A_i + B_j^{*} B_i\Bigr)\tilde{f}_{i,k}S_{i,j}\\[2mm]
        &- \sum_{i, j}^{M} \sum_{k=1}^{N} \omega_{k}\Bigl(A_j^{*} A_i + B_j^{*} B_i\Bigr)
        \Bigl(f_{i,k}f_{j,k}^{*}+\tilde{f}_{i,k}\tilde{f}_{j,k}^{*}\Bigr)
        \Bigl(\tilde{f}_{i,k}-\frac{1}{2}\tilde{f}_{j,k}\Bigr)S_{i,j}\\[2mm]
        &- \sum_{i, j}^{M} \sum_{k=1}^{N} \lambda_{k}^{'} \Bigl(A_j^{*} B_i + B_j^{*} A_i\Bigr)\Bigl[(f_{j,k}+f_{j,k}^{*})+(\tilde{f}_{j,k}+\tilde{f}_{j,k}^{*})\Bigr]
        \Bigl(\tilde{f}_{i,k}-\frac{1}{2}\tilde{f}_{j,k}\Bigr)S_{i,j}\\[2mm]
        &- \sum_{i, j}^{M} \sum_{k=1}^{N} \lambda_{k}^{'} \Bigl(A_j^{*} B_i + B_j^{*} A_i\Bigr)S_{i,j}\,.
    \end{aligned}
\end{equation}
\begin{equation}
    \begin{aligned}
        0 =\; & i \hbar \sum_{i,j}^{M} \sum_{k=1}^{N}\Bigl(A_{j}^{*}\dot{A}_{i}+B_{j}^{*}\dot{B}_{i}\Bigr)
        \Bigl(-g_{i,k}+\frac{1}{2}g_{j,k}\Bigr)S_{i,j}\\[2mm]
        & - i\hbar\sum_{i,j}^{M} \sum_{k=1}^{N}\Bigl(A_{j}^{*}A_{i}+B_{j}^{*}B_{i}\Bigr)
        \dot{g}_{i,k}S_{i,j}\\[2mm]
        &+ i\hbar\sum_{i,j}^{M} \sum_{k=1}^{N}\Bigl(A_{j}^{*}A_{i}+B_{j}^{*}B_{i}\Bigr)
        \Bigl(\dot{g}_{i,k}\,g_{j,k}^{*} - \frac{\dot{g}_{i,k}\,g_{i,k}^{*}+g_{i,k}\,\dot{g}_{i,k}^{*}}{2} + \dot{\tilde{g}}_{i,k}\,\tilde{g}_{j,k}^{*} - \frac{\dot{\tilde{g}}_{i,k}\,\tilde{g}_{i,k}^{*}+\tilde{g}_{i,k}\,\dot{\tilde{g}}_{i,k}^{*}}{2}\Bigr)\\[2mm]
        &- \sum_{i,j}^{M} \sum_{k=1}^{N}\frac{\omega_{0}}{2}\Bigl(A_{j}^{*}A_{i}-B_{j}^{*}B_{i}\Bigr)
        \Bigl(g_{i,k}-\frac{1}{2}g_{j,k}\Bigr)S_{i,j}\\[2mm]
        &- \sum_{i,j}^{M} \sum_{k=1}^{N} \omega_{k}\Bigl(A_{j}^{*}A_{i}+B_{j}^{*}B_{i}\Bigr)
        g_{i,k}S_{i,j}\\[2mm]
        &- \sum_{i,j}^{M} \sum_{k=1}^{N} \omega_{k}\Bigl(A_{j}^{*}A_{i}+B_{j}^{*}B_{i}\Bigr)
        \Bigl(g_{i,k}g_{j,k}^{*}+\tilde{g}_{i,k}\tilde{g}_{j,k}^{*}\Bigr)
        \Bigl(g_{i,k}-\frac{1}{2}g_{j,k}\Bigr)S_{i,j}\\[2mm]
        &- \sum_{i,j}^{M} \sum_{k=1}^{N} \lambda_{k}^{'}\Bigl(B_{j}^{*}A_{i}+A_{j}^{*}B_{i}\Bigr)
        \Bigl[(g_{j,k}+g_{j,k}^{*})+(\tilde{g}_{j,k}+\tilde{g}_{j,k}^{*})\Bigr]
        \Bigl(g_{i,k}-\frac{1}{2}g_{j,k}\Bigr)S_{i,j}\\[2mm]
        &- \sum_{i,j}^{M} \sum_{k=1}^{N} \lambda_{k}^{'}\Bigl(B_{j}^{*}A_{i}+A_{j}^{*}B_{i}\Bigr)S_{i,j}\,.
    \end{aligned}
\end{equation}
\begin{equation}
    \begin{aligned}
        0 =\; & i \hbar \sum_{i,j}^{M} \sum_{k=1}^{N}\Bigl(A_{j}^{*}\dot{A}_{i}+B_{j}^{*}\dot{B}_{i}\Bigr)
        \Bigl(-\tilde{g}_{i,k}+\frac{1}{2}\tilde{g}_{j,k}\Bigr)S_{i,j}\\[2mm]
        & - i\hbar\sum_{i,j}^{M} \sum_{k=1}^{N}\Bigl(A_{j}^{*}A_{i}+B_{j}^{*}B_{i}\Bigr)
        \dot{\tilde{g}}_{i,k}S_{i,j}\\[2mm]
        &+ i\hbar\sum_{i,j}^{M} \sum_{k=1}^{N}\Bigl(A_{j}^{*}A_{i}+B_{j}^{*}B_{i}\Bigr)
        \Bigl(\dot{g}_{i,k}\,g_{j,k}^{*} - \frac{\dot{g}_{i,k}\,g_{i,k}^{*}+g_{i,k}\,\dot{g}_{i,k}^{*}}{2} + \dot{\tilde{g}}_{i,k}\,\tilde{g}_{j,k}^{*} - \frac{\dot{\tilde{g}}_{i,k}\,\tilde{g}_{i,k}^{*}+\tilde{g}_{i,k}\,\dot{\tilde{g}}_{i,k}^{*}}{2}\Bigr)\\[2mm]
        &- \sum_{i,j}^{M} \sum_{k=1}^{N}\frac{\omega_{0}}{2}\Bigl(A_{j}^{*}A_{i}-B_{j}^{*}B_{i}\Bigr)
        \Bigl(\tilde{g}_{i,k}-\frac{1}{2}\tilde{g}_{j,k}\Bigr)S_{i,j}\\[2mm]
        &- \sum_{i,j}^{M} \sum_{k=1}^{N} \omega_{k}\Bigl(A_{j}^{*}A_{i}+B_{j}^{*}B_{i}\Bigr)
        \tilde{g}_{i,k}S_{i,j}\\[2mm]
        &- \sum_{i,j}^{M} \sum_{k=1}^{N} \omega_{k}\Bigl(A_{j}^{*}A_{i}+B_{j}^{*}B_{i}\Bigr)
        \Bigl(g_{i,k}g_{j,k}^{*}+\tilde{g}_{i,k}\tilde{g}_{j,k}^{*}\Bigr)
        \Bigl(\tilde{g}_{i,k}-\frac{1}{2}\tilde{g}_{j,k}\Bigr)S_{i,j}\\[2mm]
        &- \sum_{i,j}^{M} \sum_{k=1}^{N} \lambda_{k}^{'}\Bigl(A_{j}^{*}B_{i}+B_{j}^{*}A_{i}\Bigr)
        \Bigl[(g_{j,k}+g_{j,k}^{*})+(\tilde{g}_{j,k}+\tilde{g}_{j,k}^{*})\Bigr]
        \Bigl(\tilde{g}_{i,k}-\frac{1}{2}\tilde{g}_{j,k}\Bigr)S_{i,j}\\[2mm]
        &- \sum_{i,j}^{M} \sum_{k=1}^{N} \lambda_{k}^{'}\Bigl(A_{j}^{*}B_{i}+B_{j}^{*}A_{i}\Bigr)S_{i,j}\,.
    \end{aligned}
\end{equation}

\clearpage
\section{Deviation Vector}\label{Appendix C}
\begin{align}
\langle H^{2}\rangle=& \sum_{i,j}^{M}
  \Bigl\{
    \bigl(A_j^* A_i + B_j^* B_i\bigr)
    \cdot
    \Bigl(\tfrac{1}{4}\,\omega_0^2 + \sum_{k}\lambda_k^2\,\,S_{ij}\Bigr)
  \Bigr\}
\nonumber\\
&\quad
+ \omega_0 \bigl(A_j^* A_i - B_j^* B_i\bigr)
  \sum_{k}
  \Bigl(
     \omega_{lk}\,f_{jk}^*\,f_{ik}\,S_{ij}
     - \omega_{l k}\,\tilde{f}_{jk}^*\,\tilde{f}_{ik}\,S_{ij}
     + w_{r k}\,g_{ik}^*\,g_{ik}\,S_{ij}
     - \omega_{r k}\,\tilde{g}_{jk}^*\,\widetilde{g}_{ik}\,S_{ij}
  \Bigr)
\nonumber\\
&\quad
+ S_{ij}\bigl(A_j^* A_i + B_j^* B_i\bigr)
  \Bigl[
    \sum_{k}\omega_{l k}^2\,\bigl(f_{jk}^*\,f_{ik} + \tilde{f}_{jk}^*\,\tilde{f}_{ik}\bigr)
    + \sum_{k}\omega_{r k}^2\,\bigl(g_{jk}^*\,g_{ik} + \tilde{g}_{jk}^*\,\widetilde{g}_{ik}\bigr)
  \Bigr]
\nonumber\\
&\quad
+ \sum_{k,q} \,\omega_{lk}\,\omega_{l q}
  \Bigl(
     g_{jk}^*\,g_{ik}\,g_{jq}^*\,g_{iq}
     + \tilde{g}_{jk}^*\,\tilde{g}_{ik}\,g_{iq}^*\,\tilde{g}_{iq}
  \Bigr)
\nonumber\\
&\quad
- 2\sum_{k,q}\omega_{rk}\,\omega_{r q}\,f_{jk}^*\,f_{ik}\,\tilde{f}_{ji}^*\,\tilde{f}_{iq}
\quad
- 2\sum_{k,q}\omega_{r k}\,w_{r q}\,g_{ik}^*\,g_{ik}\,\tilde{g}_{ji}^*\,\tilde{g}_{iq}
\nonumber\\
&\quad
+ 2\sum_{k,q}w_{lk}\,w_{l q}\,f_{jk}^*\,f_{ik}\,g_{jk}^*\,g_{ik}
\quad
- 2\sum_{k,q}w_{rk}\,w_{r q}\,f_{ir}^*\,f_{k}\,\tilde{g}_{i\pi}^*\,\tilde{g}_{iq}
\nonumber\\
&\quad
+ \sum_{k,q}\lambda_{l k}\,\lambda_{l q}\bigl(f_{jk}^* + f_{ik}\bigr)\bigl(f_{jq}^* + f_{in}\bigr)
  \cosh\bigl(\theta_k\bigr)\,\cosh\bigl(\theta_q\bigr)
\nonumber\\
&\quad
+ \sum_{k,q}\lambda_{lk}\,\lambda_{l q}
  \bigl(\tilde{f}_{ik}^* + \tilde{f}_{ik}\bigr)\bigl(\tilde{f}_{ji}^* + \tilde{f}_{iq}\bigr)
  \sinh\bigl(\theta_k\bigr)\,\sinh\bigl(\theta_q\bigr)
\nonumber\\
&\quad
+ \sum_{k,q}\lambda_{r k}\,\lambda_{r q}\bigl(g_{ik}^* + g_{ik}\bigr)\bigl(g_{iq}^* + g_{jq}\bigr)
  \cosh\bigl(\theta_k\bigr)\,\cosh\bigl(\theta_q\bigr)
\nonumber\\
&\quad
+ \sum_{k,q}\lambda_{r k}\,\lambda_{r q}
  \bigl(\widetilde{g}_{jk} + \widetilde{g}_{jk}\bigr)\bigl(\tilde{g}_j + \tilde{g_{jq}}\bigr)
  \sinh\bigl(\theta_k\bigr)\,\sinh\bigl(\theta_k\bigr)
\nonumber\\
&\quad
- 2\sum_{k,q}\lambda_{lk}\,\lambda_{l q}\bigl(f_{jk}^* + f_{ik}\bigr)\bigl(\tilde{f}_{jq}^* +\tilde f_{iq}\bigr)
  \cosh\bigl(\theta_k\bigr)\,\sinh(\theta_q)
\nonumber\\
&\quad
- 2\sum_{k,q}\lambda_{lk}\,\lambda_{r q}
  \bigl(f_{jk}^* + f_{ik}\bigr)\bigl(\tilde{g}_{jq}^* + \tilde{g}_{iq}\bigr)
  \cosh(\theta_k)\,\sinh(\theta_q)
\nonumber\\
&\quad
+ 2\sum_{k,q}\lambda_{rk}\,\lambda_{r q}
  \bigl(f_{jk}^* + f_{ik}\bigr)\bigl(g_{jq}^* + g_{iq}\bigr)
  \cosh(\theta_k)\,\cosh(\theta_q)
\nonumber\\
&\quad
- 2\sum_{k,q}\lambda_{rk}\,\lambda_{r q}
  \bigl(\hat{f}_{jk}^* + \tilde{f}_{ik}\bigr)\bigl(g_{jq}^* + g_{iq}\bigr)
  \sinh(\theta_k)\,\cosh(\theta_q)
\nonumber\\
&\quad
+ 2\sum_{k,q}\lambda_ {rk}\,\lambda r q
  \bigl(\tilde{f}_{jk}^* + \tilde{f}_{ik}\bigr)\bigl(\tilde{g}_{ik}^* + \tilde{g}_{iq}\bigr)
  \sinh(\theta_k)\,\sinh(\theta_q)
\nonumber\\
&\quad
- 2\sum_{k,q}\lambda_{rq}\,\lambda_{rq}
  \bigl(g_{iq}^* + g_{iq}\bigr)\bigl(\tilde{g}_{iq}^* + \tilde{g}_{iq}\bigr)
  \cosh(\theta_k)\,\sinh(\theta_q)
\nonumber\\
&\quad
+ (A_i^* B_i + B_j^* A_i)\,S_{ij}
  \Bigl[
    \sum_{k}\omega_{lk}\,\lambda_{lk}\bigl(f_{jk}^* + f_{ik}\bigr)\cosh(\theta_k)
\nonumber\\
&\qquad
  - \sum_{k}\omega_{lk}\,\lambda_{lk}\bigl(\tilde{f_{jk}^*} + \tilde{f}_{ik}\bigr)\sinh(\theta_k)
\nonumber\\
&\qquad
  + \sum_{k}\omega_{r k}\,\lambda_{r k}\bigl(g_{jk}^* + g_{ik}\bigr)\cosh(\theta_k)
  - \sum_{k}\omega_{r k}\,\lambda_{r k}\bigl(\tilde{g}_{jk}^* + \tilde{g}_{jk}\bigr)\sinh(\theta_k)
\nonumber\\
&\qquad
  + \sum_{k,q} w_{lk}\,\lambda_{ l q}\,f_{jk}^*\,f_{ik}\bigl(f_{jk}^* + f_{iq}\bigr)\cosh(\theta_q)
  - \sum_{k,q}\omega_{l k}\,\lambda_{ l q}\,\tilde{f}_{jk}^*\,\tilde{f}_{ik}\bigl(\tilde{f}_{jiq}^* + \tilde{f}_{iq}\bigr)\sinh(\theta_q)
\nonumber\\
&\qquad
  + \sum_{k,q}\omega_{r k}\,\lambda_{r q}\,g_{jk}^*\,g_{ik}\bigl(g_{jk}^* + g_{iq}\bigr)\cosh(\theta_q)
  - \sum_{k,q}\omega_{r k}\,\lambda_{r q}\,\tilde{g}_{jk}^*\,\tilde{g}_{jk}\bigl(\tilde{g_{ji}} + \tilde{g}_{iq}\bigr)\sinh(\theta_q)
\nonumber\\
&\qquad
  + \sum_{k,q}\omega_{lk}\,\lambda_{r q}\,f_{jk}^*\,f_{ik}\bigl(\tilde{f}_{ji}^* + \tilde{f}_{iq}\bigr)\sinh(\theta_q)
  - \sum_{k,q}\omega_lk\,\lambda_{1q}\,\tilde{f_{jk}^*}\,\tilde{f}_{ik}\bigl(f_{jq}^* + f_{iq}\bigr)\cosh(\theta_q)
\nonumber\\
&\qquad
  + \sum_{k,q} w_{lk}\,\lambda_{r q}\,f_{jk}^*\,f_{ik}\bigl(g_{iq}^* + g_{iq}\bigr)\cosh(\theta_q)
  + \sum_{k,q}w_{r k}\,\lambda_{1 q}\,g_{ik}^*\,g_{ik}\bigl(f_{jq}^* + f_{iq}\bigr)\cosh(\theta_q)
\nonumber\\
&\qquad
  + \sum_{k,q} w_{lk}\,\lambda_{r q}\,f_{jk}^*\,f_{ik}\bigl(\tilde{g}_{jq} + \widetilde{g}_{iq}\bigr)\sinh(\theta_q)
  - \sum_{k,q}\omega_{r k}\,\lambda_{r q}\,\tilde{g}_{jk}^*\,\widetilde{g}_{ik}\,\bigl(f_{jk} + f_{iq}\bigr)\cosh(\theta_q)
\nonumber\\
&\qquad
  - \sum_{k,q}\omega_{lk}\,\lambda_{r q}\,\tilde{f}_{jk}^*\,\tilde{f}_{ik}\bigl(g_{jq}^* + g_{iq}\bigr)\cosh(\theta_q)
  \Bigr]
\nonumber\\
&\quad
+ \sum_{k,q} w_{r k}\,\lambda_{l q}\,g_{jk}^*\,g_{jk}\bigl(\tilde{f}_{jq} + \widetilde{f}_{iq}\bigr)\sinh(\theta_q)
\nonumber
\quad
- \sum_{k,q} w_{lk}\,\lambda_{r q}\,\tilde{f}_{jk}^*\,\tilde{f_{ik}}\bigl(\tilde{g}_{jq}^* + \widetilde{g}_{iq}\bigr)\sinh(\theta_q)
\nonumber\\
&\quad
+ \sum_{k,q} w_{r k}\,\lambda_{r q}\,g_{ik}^*\,g_{ik}\bigl(\tilde{g_{jq}^*} + \widetilde{g}_{iq}\bigr)\sinh(\theta_q)
\nonumber
\quad
- \sum_{k,q} w_{r k}\,\lambda_{r q}\,\tilde{g}_{jk}^*\,\tilde{g}_{i k}\bigl(g_{iq}^* + g_{iq}\bigr)\cosh(\theta_q)
\end{align}
\[
\begin{aligned}
\langle \dot{D}_M(t) \mid \dot{D}_M(t) \rangle =\; &
\sum_{l,n} \Biggl[
\Bigl( \dot{A}_l^* \dot{A}_n
   + \dot{A}_l^* A_n \sum_{k} f_{lk}^*\,\dot{f}_{nk}
   + A_l^*\,\dot{A}_n \sum_{k} \dot{f}_{lk}^*\,f_{nk} \Bigr) S_{ln}^{(f,f)}\\[2mm]
&\quad + \Bigl( \dot{B}_l^* \dot{B}_n
   + \dot{B}_l^* B_n \sum_{k} g_{lk}^*\,\dot{g}_{nk}
   + B_l^*\,\dot{B}_n \sum_{k} \dot{g}_{lk}^*\,g_{nk} \Bigr) S_{ln}^{(g,g)}\\[2mm]
&\quad + A_l^* A_n \Bigl(
    \sum_{k} \dot{f}_{lk}^*\,\dot{f}_{nk}
    + \sum_{k,q} f_{lk}^*\,\dot{f}_{nk}\,\dot{f}_{lq}^*\,f_{nq}
  \Bigr) S_{ln}^{(f,f)}\\[2mm]
&\quad + B_l^* B_n \Bigl(
    \sum_{k} \dot{g}_{lk}^*\,\dot{g}_{nk}
    + \sum_{k,q} g_{lk}^*\,\dot{g}_{nk}\,\dot{g}_{lq}^*\,g_{nq}
  \Bigr) S_{ln}^{(g,g)}
\Biggr].
\end{aligned}
\]

so that the deviation vector can be given by
\[
\begin{aligned}
\sigma^{2}(t) &= \frac{\left\|\left(i\partial_{t}-H\right)|D_{1}^{M}(t)\rangle\right\|^{2}}{\omega_{0}^{2}} \\
              &= \omega_{0}^{-2}\Bigl[\langle H^{2}\rangle - \langle \dot{D}_{2}^{M}(t) \mid \dot{D}_{2}^{M}(t)\rangle\Bigr].
\end{aligned}
\]

\twocolumngrid

\twocolumngrid

\end{document}